\documentclass[letterpaper,twocolumn,10pt]{article}
\usepackage{usenix2019_v3}

\usepackage{tikz}
\usepackage{amsmath}

\usepackage{filecontents}
\usepackage{cite}
\usepackage{amsmath,amssymb,amsfonts}
\usepackage{graphicx}
\usepackage{textcomp}
\usepackage{tabularx}
\usepackage{subcaption}
\usepackage{makecell}
\usepackage{url}
\usepackage[switch]{lineno}
\usepackage{listings}
\usepackage{multirow}

\usepackage[margin=0.5in]{geometry}
\usepackage{textcomp}

\usepackage{listings}

\colorlet{punct}{red!60!black}
\definecolor{background}{HTML}{EEEEEE}
\definecolor{delim}{RGB}{20,105,176}
\colorlet{numb}{magenta!60!black}

\begin{document}

\date{}

\title{An Open-Source P4\textsubscript{16} Compiler Backend for Reconfigurable Match-Action Table Switches}

\author{
{\rm Debobroto Das Robin}\\
Kent State University
\and
{\rm Javed I. Khan}\\
Kent State University
} 

\maketitle
\begin{abstract}

  The P4 language has become the dominant choice for programming the \textit{reconfigurable match-action table}  
  based programmable switches. V1Model architecture is the most widely available realization of this paradigm. 
  The open-source compiler frontend developed by the P4 consortium can execute syntax analysis and derive a 
  hardware-independent representation of a  program written using the latest version of P4 (also known as P4\textsubscript{16}). 
  A compiler backend is required to map this intermediate representation to the hardware resources of a V1Model switch.
  However, there is no open-source compiler backend available to 
  check the realizability of a P4\textsubscript{16} program over a V1Model   switch. 
  Proprietary tools provided by different hardware vendors are available for this purpose. 
  However, they are closed source and do not provide access to the internal mapping mechanisms. 
  Which inhibits experimenting with new mapping algorithms and innovative instruction sets for  
  \textit{reconfigurable match-action table}  architecture. 
  Moreover, the proprietary compiler backends are costly and come with various non-disclosure agreements. 
  These factors pose serious  challenges to programmable switch-related research. 
  In this work, we present an open-source P4\textsubscript{16} compiler backend for the V1Model architecture-based programmable switches.
  It uses heuristic-based mapping algorithms to map a P4\textsubscript{16} program over the hardware resources of a V1Model switch. 
  It allows developers to rapidly prototype different mapping algorithms. 
  It also gives various resource usage statistics of a P4\textsubscript{16} program, enabling comparison among 
  multiple P4\textsubscript{16} schemes. 
  
      \end{abstract}

\maketitle

\section{Introduction}

\textbf{RMT architecture and the P4 language}:
In recent years, reconfigurable match-action table (RMT)~\cite{bosshart2013forwarding} architecture-based 
programmable switches have become increasingly popular and have seen widespread 
deployment. The P4 language has emerged as the de-facto standard language to program these switches. 
Since its introduction~\cite{bosshart2014p4}, the P4 programming language has undergone several architectural changes. 
Its latest version (version 16~\cite{p416}, also known as P4\textsubscript{16})\footnote{Through the rest of the work, by P4, we mean  
P4\textsubscript{16}~\cite{p416}, unless mentioned otherwise.}
is a major redesign from the initial version (P4\textsubscript{14}~\cite{p414}) of 
the language. It  is designed to support 
various \textit{target} switches with different  architectures (i.e., software switch~\cite{shahbaz2016pisces}, 
smartNIC~\cite{piasetzky2018switch}, eBPF~\cite{tu2018linux}, FPGA~\cite{wang2017p4fpga}, RMT~\cite{opentofino}, dRMT~\cite{chole2017drmt} etc.)  
for packet processing.

RMT architecture~\cite{bosshart2013forwarding} based switches are designed as multistaged pipelines containing  
reconfigurable parser, multiple match-action stages,  deparser, and a few other fixed 
blocks (i.e., packet replication engine and traffic manager, etc.~\cite{bosshart2013forwarding,robin2020toward,opentofino}).
P4 provides language constructs to describe these switch's architecture and runtime behavior. 
The \textit{architecture description} of a switch consists of the
pipeline's high-level structure, capabilities, and interfaces. The functions supported by the hardware are 
provided as separate \textit{target-specific libraries}. 
Both of them are supplied by hardware vendors. 
The data plane program developers use the target-specific libraries and the P4 core libraries to describe the run
time behavior of an RMT switch as a P4 program.

\begin{figure}[t]
  \centering
  \includegraphics[trim=0.0in 0in 0.0in 0, clip,scale=.35]{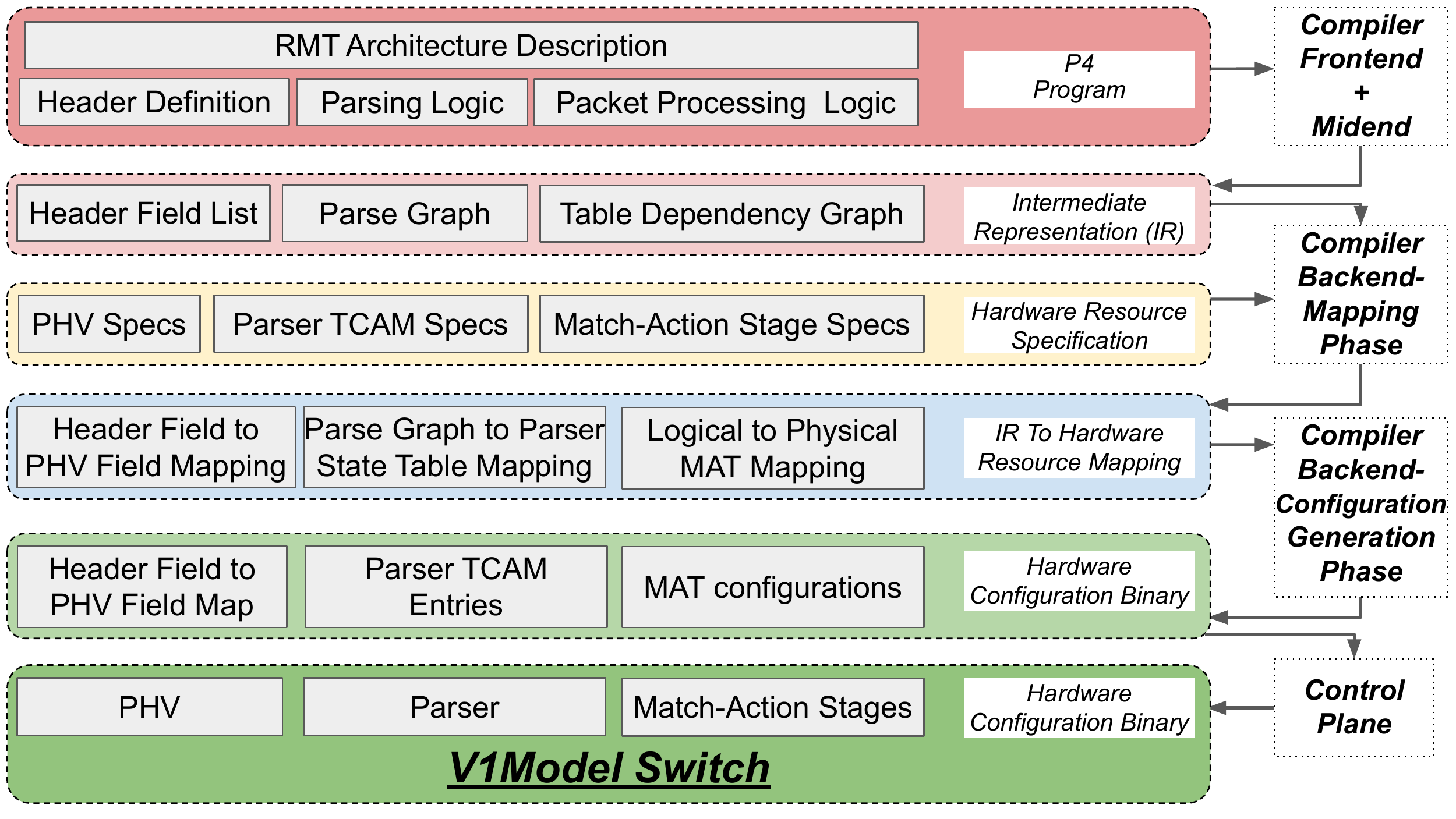}
  \caption{High level workflow of a P4 compiler for V1Model switches}
  \label{fig:CompilerArchitecture}
\end{figure}

\textbf{A P4 compiler for RMT switches}: The P4 language  is target hardware  independent in nature and provides high-level imperative  
constructs to express packet processing logics for various packet processing architectures.  
Hence, there is no direct mapping between the  P4 program
and the RMT architecture's components. A P4 compiler  is required to translate a given P4 program into a target-specific 
executable program (hardware configuration binary) to be executed by a \textit{target switch}. 
Usually, a P4 compiler (fig.~\ref{fig:CompilerArchitecture}) consists of three main components~\cite{budiu2016architecture}: 
a) a \textit{target-independent frontend} responsible for syntax analysis, 
verification of target-independent constraints (e.g., loop-free control flow required for P4) and transforming 
the P4 program into a target-independent intermediate representation (\textit{IR}) presenting the control flow between a series 
of \textit{logical match-action table}s. 
b) a midend for architecture-independent optimizations~\cite{dangeti2018p4llvm}
and 
c) a target-dependent backend responsible
for generating executable programs to be executed by the target hardware. 
It requires a resource allocation mechanism (\textit{Compiler Backend-Mapping Phase} in
fig.~\ref{fig:CompilerArchitecture}) to map the IR components on the target
hardware resources. It computes the P4 program's \textit{header field} to the RMT hardware's packet header vector (PHV) mapping,
packet header parser state machine (represented as \textit{Parse Graph} in the IR) to RMT hardware's state table mapping and 
the P4 program's control flow (represented as a graph of logical match-action tables) to RMT hardware's physical match-action table mapping.
These mappings need to
be conformant with the target-dependent constraints (i.e., header
vector capacity, crossbar width, match-action table dimensions,
etc.). If the P4 program can be successfully mapped onto the
target hardware; corresponding \textit{hardware configurations} are generated from the 
mapping (\textit{Compiler Backend-Configuration Generation Phase} in 
fig.~\ref{fig:CompilerArchitecture}) in the form of an executable hardware configuration binary.
This executable configuration is loaded into the target hardware
by the control plane and executed by the target hardware.

\textbf{Open source P4 compilers for RMT switch}: 
P4C~\cite{P4C} is the reference compiler implementation for the P4 language. 
It is developed by the  \textit{P4 Language Consortium} and 
follows~\cite{budiu2016architecture,P4C} the workflow shown in fig.~\ref{fig:CompilerArchitecture}.  
It supports two different RMT architecture based switches: 
a) simple\_switch model widely known as V1Model architecture~\cite{V1Model}
and b) Portable Switch Architecture (PSA)~\cite{PSA} developed by the  \textit{P4 Language Consortium} (not fully implemented yet). 
However, P4C  does not provide any  compiler backend for the real life target hardware of these two architectures. 
The P4C frontend+midend emits the intermediate representation as a hardware-independent JSON file, and the 
reference software switch implementation (BMV2~\cite{BMV2})  executes them over a 
CPU simulated version of the respective hardware architecture. 
It does not consider the practical hardware resource limitations that exist in real target switches.
Hence, P4C can not decide about the realizability of the given P4 program over a specific instance of these RMT switches. 
Besides the P4C, several other open-source compilers for RMT architecture-based switches are available in the literature. 
However, some of them~\cite{jose2015compiling} work with the older version (P4\textsubscript{14}~\cite{p414}) of the
P4 language, which is architecturally different from the current version of P4 (P4\textsubscript{16}~\cite{p416}). Some other works focus 
on different packet processing languages (e.g., Domino~\cite{sivaraman2016packet,gao2020switch}),
different architecture (drmt~\cite{chole2017drmt}),
or different hardware platforms (e.g., FPGA~\cite{wang2017p4fpga}). 
As a result, researchers need to use 
proprietary  compiler backends~\cite{P4Studio} to decide whether a P4 program can be implemented using an RMT switch or not. 
However, these systems are closed source, expensive, and often come with additional non-disclosure agreements~\cite{opentofino}.

\textbf{Why open-source compiler backend}: 
The compiler backend plays a crucial role in the P4 ecosystem by mapping a P4 program to the 
target hardware. It is responsible for measuring a P4 program's resource consumption in the RMT pipeline. 
Programmable switches contain a limited amount of hardware resources. Therefore, the P4 program with the least hardware resource 
required to achieve a specific task is more resource-efficient. 
In recent times, a large number of research works have 
used the BMV2~\cite{BMV2} simulator with the P4C compiler as their target platform, which lacks a 
compiler backend that can consider the real-life resource constraints present in the  target hardware. 
Without such a compiler backend, researchers can not measure the resource 
requirement of their schemes and can not compare the 
resource usage efficiency of multiple schemes. In the worst case, their P4 program may not be realizable using P4 switches, 
which are not even identifiable without a compiler backend. 
Thus, it stands to argue whether these P4 programs are directly executable over real-life RMT switches.

A compiler backend needs to 
address several computationally intractable problems~\cite{jose2015compiling,vass2020compiling}
to find the mapping of a P4 program to the target hardware. 
The optimal algorithms often require a long time to finish~\cite{jose2015compiling,vass2020compiling}. 
With the growing rise of the \textit{in-network computing}~\cite{sapio2017network} paradigm,
various research works~\cite{macdavid2021p4,robin2022clb,dang2016paxos,robin2022p4te,robin2021p4kp} are also focusing on delegating different 
\textit{network function}s in the  data plane. In these cases, researchers  
do not need to fit large P4 programs required for full-fledged switches with various features. 
Optimal mapping algorithms are useful when the data plane
programmers need to fit such large P4 programs into target
hardware. On the other hand, an open-source compiler backend that uses heuristic-based algorithms 
can  provide the researchers a quick decision about a 
smaller P4 program's realizability using a target hardware.

The mapping algorithms used in the compiler backend are 
sensitive~\cite{jose2015compiling} to the resource (TCAM/SRAM storage, number of ALUs, crossbar width, etc.) 
requirements of a P4 program and 
the available resources in a target switch. 
The resource requirements of a P4 program can change at run time (e.g., an increase in the size of an IPv4 forwarding table), 
which can invalidate a previously computed mapping. 
With the rapid proliferation of \textit{network virtualization}~\cite{hancock2016hyper4} and 
\textit{Network-as-a-Service}~\cite{zhou2010services} 
paradigm, the requirement for on-demand network function deployment is also growing rapidly. 
It requires  quick and automated deployment of customized data plane algorithms on a short notice.
Therefore, developing faster and more efficient heuristic/approximate mapping
algorithms carry enormous significance here. 
With a closed source compiler backend, researchers can not experiment with different mapping algorithms. 
Besides this, there is a growing focus~\cite{da2018extern,seibulescu2020leveraging,karrakchou2020endn} on developing 
hardware units for supporting complex instructions (\textit{extern}~\cite{p416} in P4 language) in RMT architecture. 
Without an open-source compiler backend, researchers can not integrate newly developed externs in a P4 program and test their 
effectiveness.  
Independently developing compiler backend from scratch requires various common and repetitive tasks (i.e., IR parsing, 
representing parsed IR using a graph data structure, modeling hardware resources, etc.) not directly related to 
the computation of the mappings. An open-source compiler backend can allow researchers to 
focus on developing efficient mapping algorithms rather than focusing on the repetitive tasks.


Inspired by these factors, in this work, we present the design of an open-source P4 compiler backend (mapping phase only)
for \textbf{V1Model}~\cite{V1Model} architecture-based RMT switches. 
To the best of our knowledge, it is the first open-source P4\textsubscript{16} compiler backend 
for RMT architecture based programmable switches. 
The compiler backend requires two inputs: a) a specification of the available resources in a V1Model switch and 
b) the intermediate representation (IR) of a P4 program generated by the P4C frontend. As the P4C  does not provide 
any interface to specify the hardware resources of a V1Model switch, 
we have developed JSON format based hardware specification language (HSL) (sec.~\ref{HSLSection}) to express
the \textit{hardware resource specification}s of a V1Model switch. 
After discussing the related works in sec.~\ref{RelatedWorks}, we briefly discussed 
the V1Model architecture in sec.~\ref{V1ModelArchitecture} along with the HSL (sec.~\ref{HSLSection}).
Then we present the structure of the IR provided by the P4C compiler frontend (sec.~\ref{IR}).
This backend uses various existing heuristic-based algorithms to allocate resources in the V1Model switch pipeline and 
computes  the \textit{IR to hardware} resource mapping. 
To the best of our knowledge, this is the first scheme in literature which considers the constraints
arising from the use of  stateful memory in a P4 program and its impact on the mapping decision. 
We discussed the details of the mapping process in sec.~\ref{MappingProblem}.
Once the mapping is found, computing the hardware configuration binaries 
requires a straightforward translation of  the mapping into hardware instruction  codes. 
As this work does not focus on 
executing a P4 program on any specific instance of V1Model switch, 
we leave the hardware configuration binary generation for future work. 
We discuss  the implementation and evaluation of our compiler backend in sec.~\ref{ImplementationAndEvaluation} 
and conclude the paper in sec.~\ref{Conclusion}.

\textbf{Why V1Model}: 
V1Model is the only RMT switch fully supported by the open-source P4C compiler frontend. Moreover, V1Model architecture is supported 
by major programmable switch hardware vendors~\cite{harkous2020p8,opentofino}. In recent years, a large set of research works~\cite{hauser2021survey} have used 
the V1Model as their reference hardware architecture (either through the use of commercial hardware or a BMV2 simulator). 
Moreover, V1Model is similar to the abstract switch model used in the P4 language version 14. Therefore, all the P4\textsubscript{14} 
based  research works can also be mapped to this model. Finally, the latest programmable switch architecture being standardized by 
the P4 consortium is PSA~\cite{PSA}, and it is also similar to the  V1Model architecture. The compiler backend presented in this 
work can be extended for PSA architecture with a small number of modifications.
Due to these reasons, V1Model is a representative hardware architecture for a large number of research works, and we have chosen to 
build the compiler backend for this architecture.

\textbf{What the compiler-backend is not}: The compiler backend presented in this work supports only V1Model architecture and 
a subset of P4 language (P4\textsubscript{16}) constructs which cover a wide range of use cases.
A full list of P4 constructs supported by the system can be found at~\cite{P4CB}. 
Proprietary hardware can have special instructions (as \textit{extern}~\cite{p416}) available for packet processing, 
and they can also have additional
constraints in their system. Our compiler backend is not a full replacement of any proprietary system. 
It uses heuristic algorithms for mapping a P4 program to V1Model switches, and it can sporadically reject a P4 program 
(like other programmable switch backends~\cite{gao2019autogenerating}) despite the existence of a valid mapping. 
Moreover, due to the use of heuristics, it does not guarantee the optimality of the computed mapping. 
Finally, the compiler backend only covers the mapping phase shown in fig.~\ref{fig:CompilerArchitecture} 
and does not cover the \textit{hardware configuration generation} phase.

\section{Related Works} \label{RelatedWorks} 

In~\cite{bosshart2014p4}, the authors introduced an RMT architecture based \textit{abstract switch forwarding model}
and presented the P4 programming language to program them in a protocol-independent manner. 
The authors also presented the high-level structure of a two-stage P4 language compiler.
Though the work briefly discussed the parser and TDG mapping problem, a complete open-source system for the 
compiler backend was absent. 
In~\cite{gibb2013design}, the authors addressed the problem of mapping a packet parsing logic to CAM-based hardware. 
However, its main focus was on synthesizing parser hardware circuitry. Hence it can not be 
directly used in a P4\textsubscript{16} compiler backend.
In~\cite{jose2015compiling}, the authors discussed the computational complexity of  mapping the logical match-action tables 
to the physical match-action tables of an RMT switch.
They presented an integer linear programming-based method (for optimal solution) and 
a few heuristic-based methods for computing the mapping. The system is available as an open-source project. However, 
it can not support  stateful memories in the P4 program,
which is a crucial requirement for the in-network computation paradigm.
All the works mentioned above were designed to support the initial version of the P4 language (a.k.a. P4\textsubscript{14}~\cite{p414}),
and none of them provides a complete compiler backend. 
Moreover, the latest version of the P4 language (a.k.a. P4\textsubscript{16}) is architecturally different 
from the P4\textsubscript{14}. 
Hence these works can not be used directly for compiling P4\textsubscript{16} programs.

The reference compiler for the P4\textsubscript{16} language developed by the P4 language consortium is P4C~\cite{P4C}.
Its frontend can compile a P4\textsubscript{16} program for  various target architectures (including the RMT architecture). 
It provides backend support through CPU-based simulation for two RMT architecture switches: V1Model~\cite{V1Model} and PSA~\cite{PSA}.
These simulated backends execute the \textit{intermediate representation} of a P4 program over a CPU. 
It has no provision to model the hardware resources of an RMT switch. 
It is unable to consider the hardware resource limit constraints in deciding whether a P4\textsubscript{16} program can be 
mapped to target hardware in real life or not. 
In~\cite{wang2017p4fpga}, the authors presented an open-source P4\textsubscript{16} compiler backend for FPGA-based platforms. 
However, this system's basic blocks are different from the physical match-action tables used in RMT architecture. 
Here, the basic blocks can execute both match, and branching instructions; and based on their result, some actions can be executed. 
Hence, it provides a more flexible match-action capability 
in every node compared to the original RMT architecture. 

Few other open-source compilers compiler~\cite{gao2019autogenerating,voros2018t4p4s} exist
in the literature for the RMT architecture. 
However, they~\cite{gao2019autogenerating} either do not
support the program written in P4\textsubscript{16} as the input or are designed
for non-RMT architecture-based hardware platforms~\cite{voros2018t4p4s}. 
Besides
these open-source systems, a few proprietary compiler backends~\cite{gao2020lyra,P4Studio}
capable of supporting the P4\textsubscript{16}  language for RMT switches also exist. However,
they are  closed source in nature and do not provide access to their internal mechanisms.

\section{V1Model Architecture  } \label{V1ModelArchitecture} 
\begin{figure}[b]
 \centering
 \includegraphics[trim=0in 1in 0in 0, clip,scale=.345]{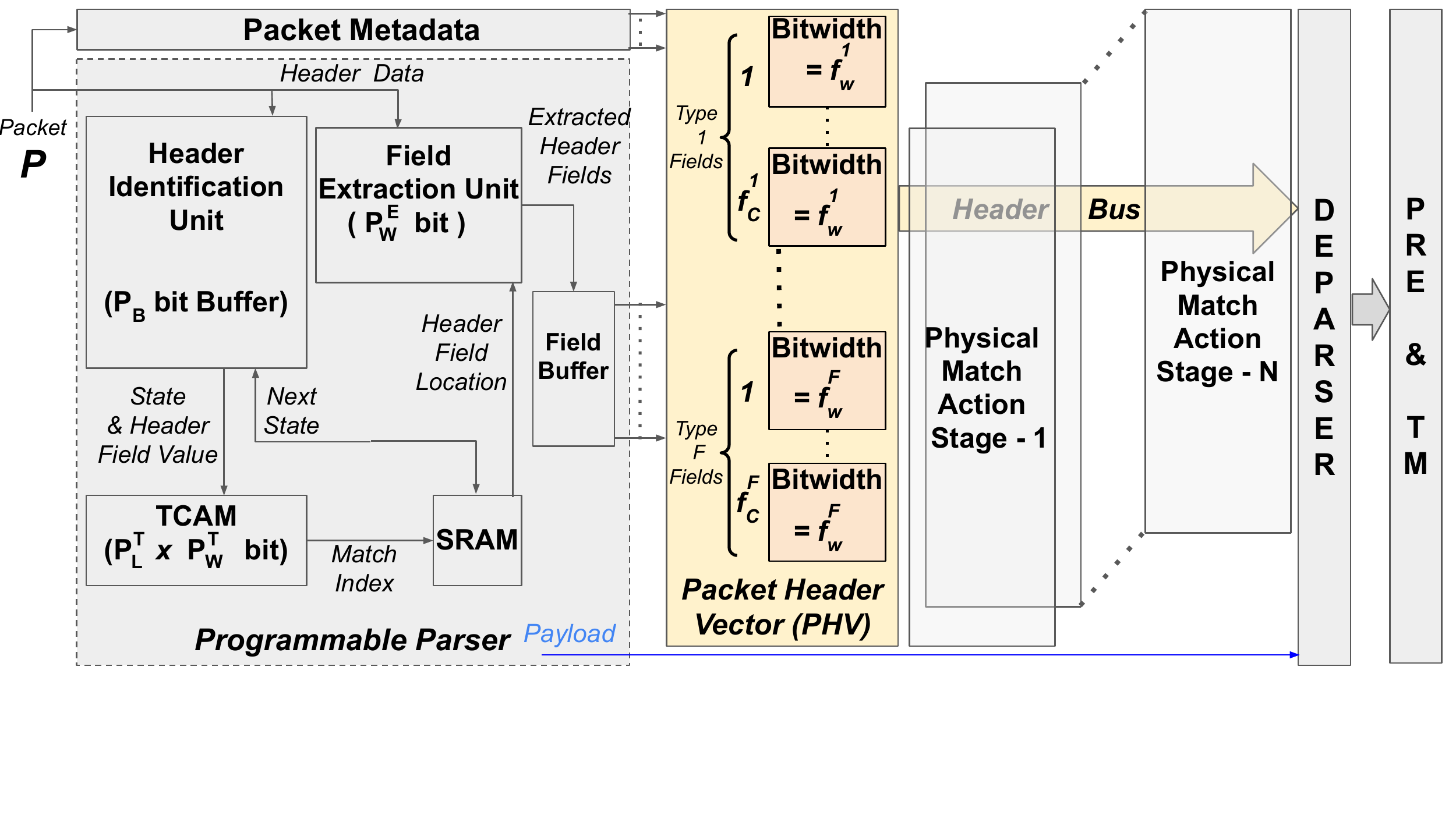}
 \caption{ V1Model pipeline architecture}
 \label{fig:V1ModelArchitecture}
\end{figure}

The V1Model   is an instance of a
\textit{reconfigurable match-action} (RMT) architecture. 
Its packet processing pipeline (fig.~\ref{fig:V1ModelArchitecture}) consists of several components arranged in multiple stages. 
This section describes its components, the specification of 
different resource types, and how they process a packet. Finally, in sec.~\ref{HSLSection} we present a hardware 
specification language to represent a V1Model switch's resources.

\subsection{Parser and Packet Header Vector} \label{ParserPacketHeaderVector}
In V1Model architecture, an incoming packet at first goes through a TCAM based~\cite{gibb2013design} 
\textit{programmable parser} (fig.~\ref{fig:RMTSingleStage}), which executes the parsing logic provided 
in the form of a  state machine (converted to a \textit{state table} 
by a compiler backend). 
The parser contains two main building blocks: 

a) \textit{Header Identification Unit}: 
It contains a $P_B$ bit wide buffer to look ahead in the packet and identify maximum $H$ headers every cycle. 
It also contains a  
TCAM capable of storing $P_L^{T}$ entries to implement the \textit{state table}. Every TCAM entry contains information of a 
current parsing state and values (as bit sequence) of header fields to be matched. 
At every cycle, maximum $f_C^T$ lookup field values (each having maximum lookup width $f_W^T$ b)   
and the \textit{current state} can be looked up in the TCAM. The TCAM entries are $P_W^T$ b wide to
store the lookup field values and the current state value.
Every entry also contains a 
pointer to RAM cells  for storing the \textit{next parsing state} and the  location of the header fields 
to be extracted by the \textit{extraction unit}.

b) \textit{Extraction Unit}: 
After matching a packet in the TCAM, the information stored at the \textit{match index} th cell of the SRAM is loaded into 
the \textit{extraction unit}. This unit can extract maximum $P_W^{E}$ bit wide data as header fields 
and store them in a \textit{field buffer}.
At every cycle, a few header fields are extracted, and 
the \textit{next parsing state} is fed to the \textit{header identification} unit for matching in the  TCAM in the next cycle.
The header identification unit can move ahead to a maximum $P_{MA}$ bit  in the packet to start identifying the next header fields. 
Every parser unit is designed for a maximum parsing rate ($P_{Rate}$) throughput. 
V1Model switches can deploy multiple parser units parallelly for achieving a higher packet parsing rate. 

After completing the parsing, all the extracted header fields  are sent to 
a \textit{Packet Header Vector} (PHV) from the \textit{field buffer}. 
The PHV can store $F$ different types of  fields;  all ${f}_C^i$ header fields of  type $i$ are ${f}_{W}^i$ bit wide ($i = 1$ to $F$ ).
Multiple fields in the PHV can be merged together to form a larger header field. 
Besides the parsed header fields, a PHV also stores hardware-specific metadata (i.e., ingress port, timestamp, etc.).
The PHV is passed to subsequent components ($N$ match-action stages fig.~\ref{fig:V1ModelArchitecture}) 
in the pipeline through a wide \textit{header bus}.

\subsection{Match-Action Stages} \label{MatchActionStagesSubSection}

\begin{figure}[b]
 \centering
 \includegraphics[trim=0.1in 0in 0in 0, clip,scale=.35]{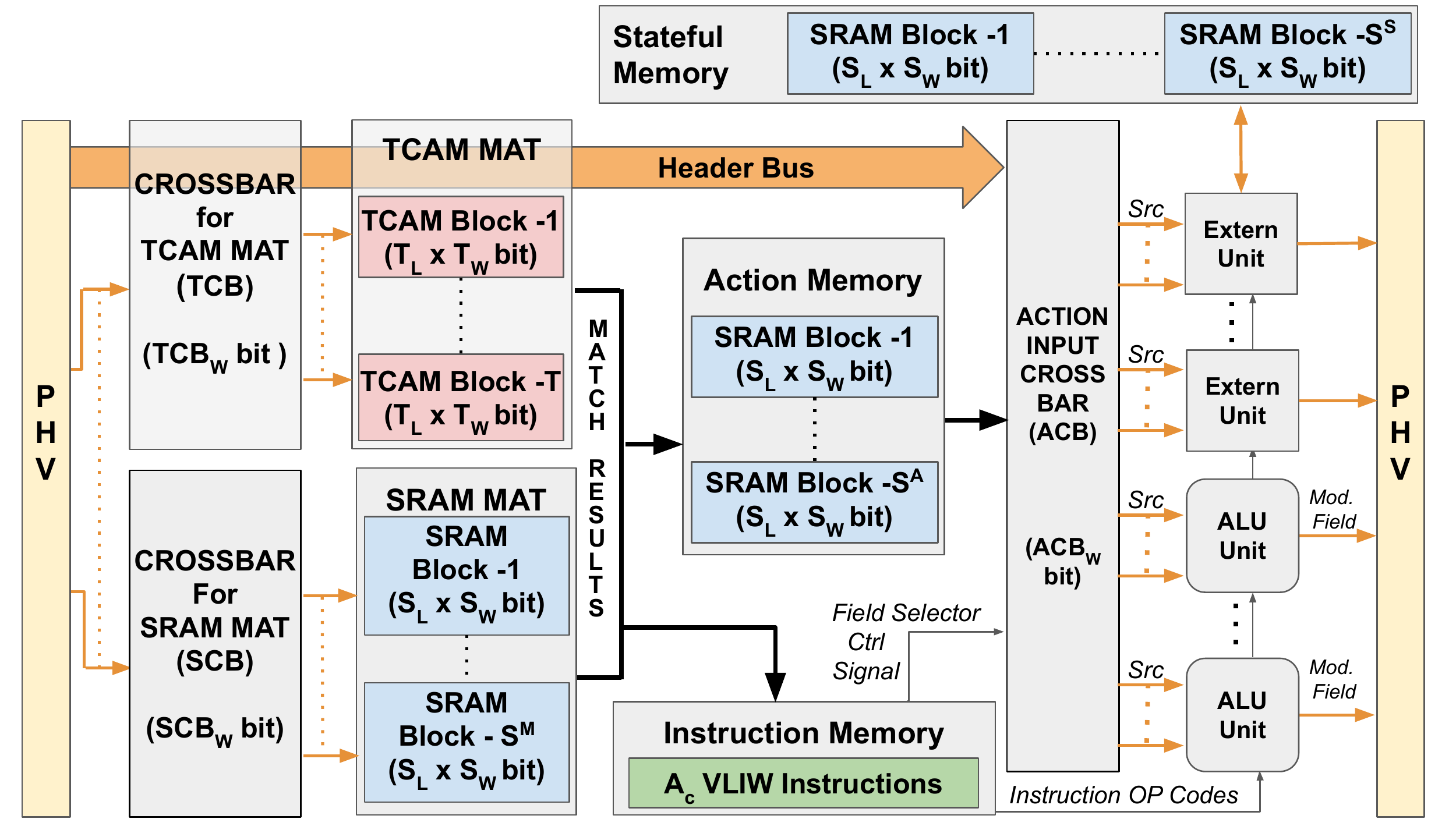}
 \caption{ A match-action stage in RMT pipeline}
 \label{fig:RMTSingleStage}
\end{figure}

Next, the PHV goes through a series of $N$ match-action stages for \textit{ingress stage} processing. 
Each stage (fig.~\ref{fig:RMTSingleStage}) contains $T$ units of $T_{W}$ bit wide  TCAM blocks, each capable of storing  $T_{L}$ entries. 
It also contains $S$ units of $S_{W}$ bit wide SRAM blocks, each capable of storing $S_{L}$ entries. 
The TCAM blocks are used to implement   \textit{physical match-action table}s (MAT) for ternary/range/prefix/exact matching. 
A fraction of the SRAM blocks ($S^M$ blocks) 
are used to implement hashtable (using $HS_K$ way Cuckoo hashtable~\cite{pagh2004cuckoo,kirsch2010more})  
based \textit{physical match-action table}s   
for exact matching, and the rest 
are used for storing other information (i.e., action arguments, next MAT address, etc.). 
These smaller \textit{physical match-action tables} can be run independently or  grouped 
together to match wider header fields within a stage. 
Physical MATs across the stages can be merged to implement a longer table. 
Header fields are supplied from PHV to the TCAM and SRAM based physical MATs through two crossbars, TCB (${TCB}_W$ bit wide) 
and SCB (${SCB}_W$  bit wide), respectively. 
With every entry in the MATs, there is a pointer to corresponding action information (action arguments, action instruction, 
address of the next MAT to be executed, etc.).  On finding a match in the MATs, the corresponding action 
information is loaded from the memory. Every match-action stage contains a separate arithmetic logic  unit (ALU) for every field of the PHV for 
parallel computation. Two or more units can be grouped together to execute computation on larger  fields. 
Besides the per header field ALU units, a fixed number of \textit{extern} units (hash, counter, register, meter, etc.) 
are also available in every match-action stage for special operations (i.e., hash computations, counting, storing/loading states, etc.).

Every stage can store ${A}_{C}$ VLIW instructions for all the physical MATs. Every VLIW instruction carries separate instruction for  
the per header field ALU and extern units. Data is provided to these processing units from PHV through 
an ${ACB}_W$ bit wide crossbar (ACB). 
Similar to the match crossbars (TCB and SCB), every bit of this crossbar is driven by all the fields from PHV.
The action information (except the action instructions are stored in a 
dedicated memory) and the stateful memories used by the \textit{extern} units  
are allocated in separate chunks of the available SRAM blocks ($S^A$ blocks and $S^S$ SRAM blocks, respectively). 
Every stage contains $M_{P}$   memory ports (each one $M_{BW}$ bit wide) capable of read-write 
from/to an SRAM cell in one clock cycle. 
These ports are used to read/write data from SRAM blocks for exact MATs, action memory, and stateful memories.
Every TCAM based MAT can store a fixed number of match entries (up to their full capacity). On the other hand,
an SRAM-based MAT can store a variable number of entries because the same SRAM blocks are 
allocated to store match entries, action entries, and stateful memories. 
The number of total SRAM blocks ($S^M$, $S^A$, and $S^S$ out  of total $S$ blocks available) used for 
exact match MAT, action memory, and stateful memory 
depends on the \textit{logical to physical MAT mapping} algorithms of sec.~\ref{sec:TDGMapping}.
To optimize the SRAM usage, RMT architecture allows \textit{word packing}, creating a \textit{packing unit} 
of multiple SRAM blocks. Multiple entries (match, action, or stateful memory entry) can be stored in one unit to reduce 
SRAM waste. This \textit{variable packing} format does not impact the match performance, and the match units can 
match a packet against multiple words stored in the same SRAM block. 

\textbf{Packet Replication Engine and Traffic manager (PRE \& TM)}: After finishing the ingress stage processing, 
the packet is submitted to the egress port's queue. 
The \textit{PRE \& TM} is a non-programmable component responsible for handling the packet's life cycle in 
the port's queues, scheduling 
the packets, and replicating the packets if necessary. 
Besides these, there are two more fixed-function components for computing and verifying the checksum of a packet. 
As they are fixed-function blocks, we do not discuss their details. 

\textbf{Egress stage}: Once the packet is picked from the egress port's queue, it undergoes the egress stage processing. 
The egress stage is similar to the \textit{ingress stage} and shares the same physical components for their processing. 
The compiler backend allocates the resources between the ingress and egress thread in such a way that they do not 
hamper each other's packet processing activities. 

\textbf{Deparser}: After egress stage processing is finished, a packet goes through a deparser block. It 
recombines the data from the packet header vector fields and the payload. 
Then the packet is finally propagated through the outgoing channels.

\subsection{V1Model Hardware  Specification Language}\label{HSLSection}
A compiler backend requires information about the available resources of a V1Model switch.
However, the openly available P4C compiler does not provide any interface to model it. 
Packet header vector, programmable parser, and match-action stages are the major programmable components in the V1Model architecture. 
We developed a JSON format-based hardware specification language (\textit{HSL}) to specify the available resources in the programmable 
components of a V1Model architecture based switch. The language allows specifying how many header fields 
can be accommodated in a PHV and what are the bitwidth of these fields. Similarly, it allows specifying the dimension of 
various hardware resources used in the programmable parser (sec.~\ref{ParserPacketHeaderVector}). 
It also allows specifying the number of match-action stages and the number of resources in every stage as described in 
sec.~\ref{MatchActionStagesSubSection}. Appendix~\ref{APP:ExampleV1ModelSpecs} shows an example hardware specification 
of a V1Model switch.




\section{P4C Intermediate Representation} \label{IR}
The P4C frontend takes the \textit{architecture description} of the V1Model architecture and a P4 program 
as the input. 
The intermediate representation generated by the P4C frontend (along with the midend) is a 
\textit{target-independent representation} (IR) of the
P4 program. The JSON representation of the IR contains the following  major components:

\subsection{Header Information}\label{IR:Header}
The P4 language provides language constructs to represent a packet header in an object-oriented style where a header object can contain 
multiple fields. 
The IR contains a list of all the headers (including the packet metadata header) used in the P4 program
along with their member fields and their bitwidth. 

\subsection{Parse Graph}\label{IR:ParseGraph}
The \textit{parse-graph} is a directed acyclic graph representation of the parser state machine (parsing logic given in the P4 program).
It defines the sequence of headers inside a packet. 
Every node in the \textit{parse graph} represents a header type, and the edges represent a transition of the parser state machine. 
An edge from node $a$ to node $b$  indicates that after parsing header $a$, based on a specific value of one of its member header fields 
$b$ will be parsed.

\subsection{Table Dependency Graph}\label{IR:TDG}
Processing logic for \textit{ingress} and \textit{egress} stages is written using imperative programming constructs given by 
the P4  programming language. The P4C frontend converts these logics into a flow of \textit{logical match-action tables} and 
their \textit{dependencies}. 
Thus each stage's control flow is converted into a  \textit{Table Dependency Graph} (TDG).
A TDG is a directed acyclic graph where every node represents a logical MAT, and every edge represents the dependency between any two logical MATs.
Each node describes the set of fields (header and/or metadata fields) to be matched against the table entries,
the types of matching (exact, prefix, ternary, etc.), and the maximum number of entries to be stored in the memory for this table. 
It also describes the set of actions to be executed based on the match result and the address of the next table to be loaded after 
executing the current table. 

\textbf{Non-stateful memory dependecy}: Every path in the TDG represents a chain of logical MATs. Following four types of 
dependencies can arise between any two  logical MATs (\textit{A} comes first and then \textit{B}) in the same path. 
These dependencies do not involve access to any stateful memory used in the P4 program.

\hspace{2mm} \textbf{1. Match dependency}: A field is modified by the actions of  \textit{A}, and it is a match field of \textit{B}.
Hence execution of B's match part must start after table A's action part has finished execution. 

\hspace{2mm} \textbf{2. Action dependency}: A field modified by \textit{A} is not matched
in \textit{B}. The same field is also modified by \textit{B}. The modification
done by the action of later table \textit{B} becomes the final result. Hence
execution of \textit{B}'s action part must start after table \textit{A}'s action part
has finished execution.

\hspace{2mm} \textbf{3. Successor dependency}: Whether table \textit{B} will be executed or not is decided by the match result 
(hit or miss) of table \textit{A}. 
Hence execution of \textit{B}'s match part must start after table \textit{A}'s match part has finished execution. 

\hspace{2mm} \textbf{4. Reverse match dependency}: A match field of table \textit{A} is modified by the action part of table \textit{B}. 
Hence execution of \textit{B}'s action part must start after table \textit{A}'s match part has finished execution.

\textbf{Stateful memory dependency}: Another type of dependency arises between two logical MATs on the same/different paths if they access
the same set of \textit{stateful memories} (counter, meter, register, etc.). 
The P4 language provides two different ways for accessing stateful memories:
a) \textit{direct}: these stateful memories are attached to a logical MAT and do not create any dependency on other 
logical MATs. And b) \textit{indirect}: they are not directly attached to any specific MAT, and 
multiple MAT can access them. Hence they create a dependency among the tables accessing them. 
The SRAM cells are 
attached to only one match-action stage in the RMT pipeline. Hence, 
if a set of SRAM blocks in stage X of the RMT pipelines are allocated for an indirect stateful memory;
two or more logical MAT accessing it must need to be mapped on the same match-action stage X.
We term this as \textbf{stateful memory} dependency.



These dependencies influence how the logical MATs in a TDG will be mapped on the physical MATs of an RMT pipeline. 
The predecessor node in every type of dependency guides the starting clock cycle of its successor~\cite{jose2015compiling,bosshart2013forwarding}.
Hence, they also determine the total processing delay of a packet undergoing the processing defined by a P4 program. 
However, the TDG in the IR itself has no mechanism to represent various types of dependencies directly. 
A compiler backend needs to analyze the TDG to identify the dependency between two logical MAT before computing the mapping.


\section{IR to Hardware Resource Mapping} \label{MappingProblem}

The goal of the \textit{mapping phase}  shown in fig~\ref{fig:CompilerArchitecture} is to map 
a given P4 program to the resources of a  V1Model switch.  A compiler backend takes two types of information as input for this purpose: 
a) the hardware resource specification of the switch described using the HSL presented in sec.~\ref{HSLSection}
and b) the hardware-independent intermediate representation (IR) (sec.~\ref{IR}) of the given P4 program generated by the 
P4C compiler frontend. 
To determine whether the given P4 program is realizable over a V1Model switch,
the compiler backend needs to address three different types of mapping problems: 
a) header mapping: mapping the IR's header fields to the PHV fields 
b) parse graph mapping: mapping the parse graph to parser hardware resources, and 
c) TDG mapping: mapping of the TDG to the resources of match-action stages. 
While addressing these problems, the compiler backend needs to ensure that a computed mapping does not violate the 
constraints (both architectural and 
resource constraints) of the target hardware and the control flow of the P4 program. 
Besides computing a valid mapping, the backend also needs to maximize the concurrency and resource usage efficiency of the P4 program. 

Computing an \textit{optimal and valid mapping} of a P4 program to a V1Model switch 
is a computationally intractable problem~\cite{jose2015compiling,vass2020compiling}. 
Our compiler backend uses heuristic-based algorithms to find a mapping while maintaining the architectural and resource constraints 
of an RMT switch.  We describe how our compiler backend works to find these three types of mapping through the rest of this section
using a simple \textit{QoS-modifier}  program (see appendix~\ref{App:QoSModiferP4Program} for the P4 source code). In this program, 
the control plane supplies a separate QoS value for  IPv4 and IPv6 packets forwarded through an
individual port.
The P4 program stores these values in two indirect stateful memories (register array) on receiving the 
control packet from the control plane (defined in MAT named \textit{match\_control\_packet}).
For every valid IPv4 packet, if it matches with the entries configured in \textit{ipv4\_nexthop}, 
that packet's \textit{diffserv} value is replaced with a value read from the register array (\textit{ipv4\_port\_qos}),
and it's IPv6 destination is set as a server's IP address. Then the packet is matched with \textit{ipv6\_nexthop} MAT to 
set its IPv6 \textit{trafficClass} after reading from another register array (\textit{ipv6\_port\_qos}). 
In the example, we have used the benchmark V1Model switch specified in appendix~\ref{APP:ExampleV1ModelSpecs}.

\subsection{Header   Mapping} \label{sec:HeaderMapping}
V1Model switches contain a limited number of fixed bitwidth fields in the PHV (provided as a part of the hardware specification).   
The header fields used in the program is needed to be accommodated using these PHV fields to execute a P4 program.
For example, a 17b wide header field can be accommodated using three 8b or 32b wide PHV fields. 
In the first case, the amount of resource \textit{waste} is 7b and 15b in the latter case. 
Here, the mapping algorithm also needs to minimize the amount of waste.
Besides this, as the V1Model switches contain two different processing threads (for the ingress and egress stage) 
sharing the same hardware pipeline, every PHV field needs to be 
explicitly allocated to one of them. 

The problem of optimally allocating the PHV fields (set of items) to accommodate the P4 program's header fields (set of resources) 
can be modeled as a \textit{multiple knapsack} problem~\cite{puchinger2010multidimensional}. 
In this case, we need to find how to optimally fill up a set of bins (the header fields in the P4 program) using the set of items 
(all the PHV fields) while minimizing the waste. 
As the problem is computationally intractable, we used a heuristic-based 
(fill the largest header field at first using the largest bitwidth PHV field
available)
algorithm to compute the mapping. 
The algorithm first analyzes the header information in the IR (sec~\ref{IR:Header}) and 
makes two disjoint sets of header fields  used in the ingress and egress stage. 
The metadata fields are needed to be replicated for both stages. 
The header fields are sorted in descending order of their bitwidth.
For each header field in sorted order, the algorithm picks the largest bitwdith PHV field from the remaining PHV field 
that leads to the least amount of waste. The process continues until the bitwidth of a header field is filled with 
selected PHV fields. Table~\ref{tab:PhvUsageOfQosModifer} shows the bitwidth of the header fields used in 
the QoS-Modifier program (appendix~\ref{App:QoSModiferP4Program}). 
It also shows the bitwidith of the PHV fields, and how many of them are required to accommodate those header 
fields according to the mappings computed by the compiler backend.

\begin{table}
\centering
\begin{tabular}{|c|c|c|c|c|c|} 
\hline
 \multicolumn{4}{|c|}{\textbf{P4 Program Header Fields}} & \multicolumn{2}{c|}{\textbf{PHV Fields}}   \\ 
\hline
Bitwidth & \multicolumn{1}{l|}{Count} & \multicolumn{1}{l|}{Bitwidth} & Count             & Bitwidth            & Count                \\ 
\hline
1        & 1                          & 13                            & 1                 & \multirow{3}{*}{8}  & \multirow{3}{*}{49}  \\ 
\cline{1-4}
2        & 1                          & 16                            & 6                 &                     &                      \\ 
\cline{1-4}
3        & 3                          & 19                            & 2                 &                     &                      \\ 
\hline
4        & 3                          & 20                            & 1                 & \multirow{2}{*}{16} & \multirow{2}{*}{17}  \\ 
\cline{1-4}
7        & 4                          & 32                            & 7                 &                     &                      \\ 
\hline
8        & 11                         & 48                            & 2                 & \multirow{2}{*}{32} & \multirow{2}{*}{24}  \\ 
\cline{1-4}
9        & 5                          & 128                           & 2                 &                     &                      \\
\hline
\end{tabular}
\caption{Bitwidth of  \textit{QoS-Modifier} program's (appendix~\ref{App:QoSModiferP4Program}) header fields  and bitwidth of PHV fields required to accomodate them  }
\label{tab:PhvUsageOfQosModifer}
\end{table}

\subsection{Parse Graph Mapping} \label{sec:ParseGraphMapping}
The IR gives a graph (\textit{parse graph} in sec.~\ref{IR:ParseGraph}) representation of the parser state machine. On the other hand,
the parser in the V1Model switches follows \textit{match-action} semantics. 
In every cycle, the parser hardware can look into   the packet header, identify a limited number of 
header fields, and \textit{match} them 
with the header-identification unit's TCAM entries (works as a state table). On finding a match in the TCAM, 
the corresponding  \textit{action} (header fields extraction by the \textit{extraction-unit}) is executed. 
The compiler backend needs to convert 
the \textit{parse graph} into a state table which needs to be accommodatable within the dimensions of the parser TCAM. 
Similarly, the field extraction operations 
must be accommodatable within the capacities of the \textit{extraction unit}.

Generating an optimal state table from a parse graph and relevant TCAM entries is  a computationally intractable problem~\cite{gibb2013design}.
Our compiler backend uses the algorithm presented in~\cite{gibb2013design}. This algorithm tries to find the clusters in the 
\textit{parse-graph} which can be accommodated within the capacity (lookup in the packet header and extracting 
header fields) of the parser hardware. 
For every unique pair of a cluster and an edge (transition in the parser state machine) in the 
cluster graph, one entry is required in the TCAM based parse-table. 
The parser hardware executes the match-action logic for every cluster in one cycle and moves to the \textit{next parsing state}. 
We converted the source code 
of~\cite{gibb2013design}\footnote{In~\cite{gibb2013design} the authors also proposed 
several optimizations to reduce the number of clusters. We leave integrating them as future work. }
to work with the IR generated by P4C and used it in the compiler backend to 
generate the \textit{state-table} entries. 
If the entries can not be accommodated using the TCAM, then the P4 program is rejected. 
As an example, the compiler backend was used to compute the \textit{parse-graph mapping} for the 
P4 program shown in appendix~\ref{App:QoSModiferP4Program} over the V1Model hardware described in appendix~\ref{APP:ExampleV1ModelSpecs}.
Fig.~\ref{fig:ParseGraphMappingExample} shows the corresponding
\textit{parse graph}, identified clusters in it, and the corresponding TCAM entries. 

\begin{figure}[]
\centering
\includegraphics[trim=0.0in 0in 0in 0, clip,scale=.34]{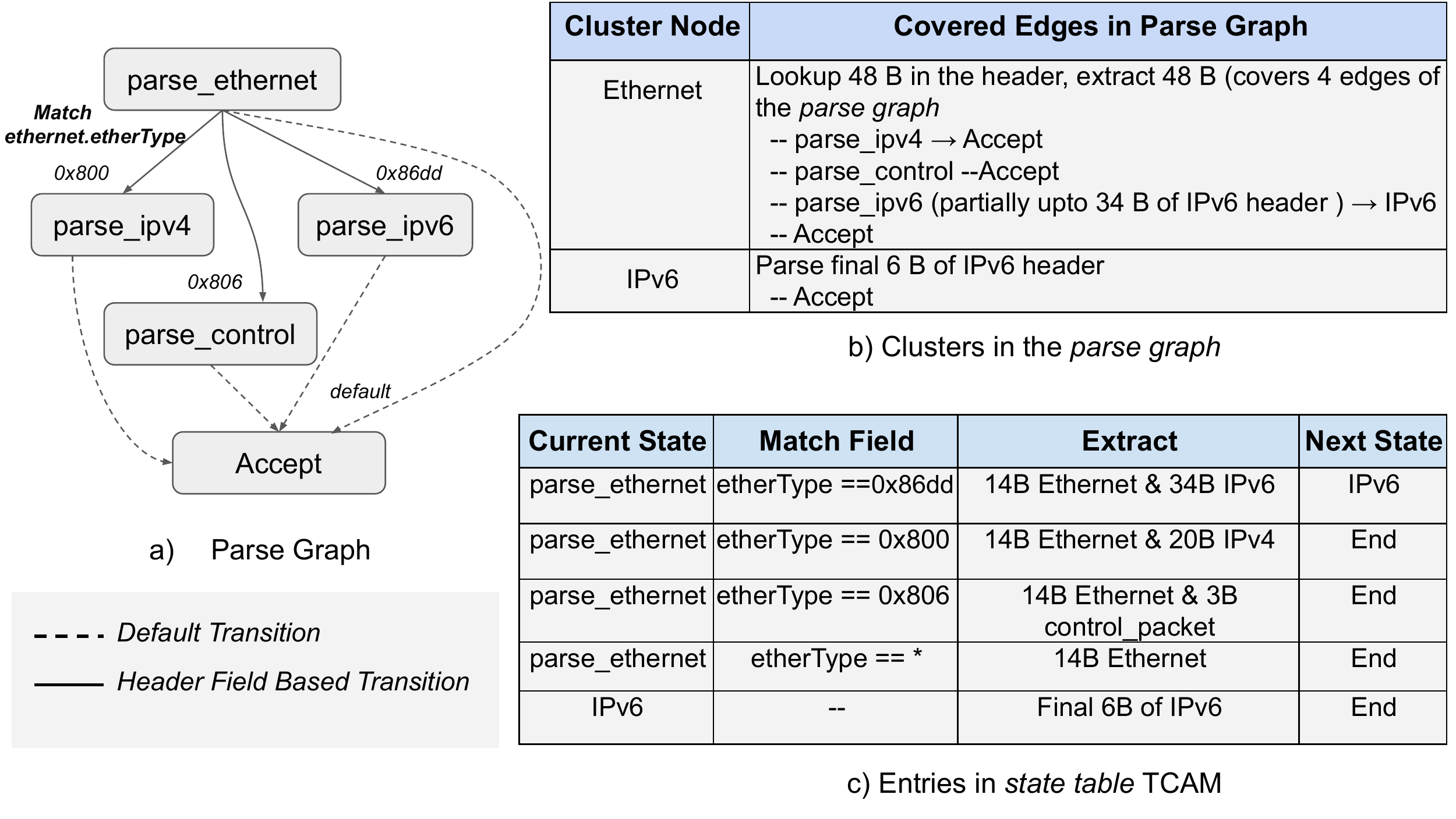}
\caption{The \textit{parse graph}, identified clusters in it, and the corresponding TCAM entries in \textit{state table} for the 
P4 program of fig.~\ref{App:QoSModiferP4Program}}
\label{fig:ParseGraphMappingExample}
\end{figure}

\subsection{TDG Mapping}\label{sec:TDGMapping}
The TDG expresses the packet processing logic of a P4 program as a control flow among logical match-action tables guided by their dependencies. 
The compiler backend needs to map these logical MATs over the physical MATs of the V1Model switch while preserving the original control flow. 
The computed mapping must not violate the following constraints.
\begin{itemize}
\item The \textit{match width}  of all the logical MATs mapped to any match-action stage must not exceed the 
crossbar width for TCAM based MATs ($TCB_W$) and SRAM based MATs ($SCB_W$).
\item The bitwidth of the match fields of every logical MAT must not exceed the total match width of the mapped physical MATs.
\item The total number of entries required for any logical MAT must not exceed the capacity of the mapped physical MATs. 
\item The total width of the fields used as the action inputs must not exceed the width of the action crossbar of a match-action stage. 
\item The total number of actions (using ALU or extern units) executed on an individual PHV field $f$ in a match-action stage 
      must not exceed the available number of 
      actions for $f$ in that stage. 
\item The total amount of SRAM required by logical MATs for action memories and 
      stateful memories (using a register, meter, counter, etc.) must not exceed 
      the available SRAM volume in a match action stage. 
\item If two or more logical MAT nodes access the same indirect stateful memory they need to be mapped on
      the same physical match-action stage.
\item The physical match action tables in a single stage can be executed concurrently. However, due to 
      the dependencies (sec.~\ref{IR:TDG}), the compiler backend needs to map the logical MATs over the physical 
      MATs in such a way that does not violate the dependencies. 
\end{itemize}
Besides ensuring the constraints, the compiler backend needs to minimize the number of match-action 
stages and amount of resource usage in each of the used match-action stages. The first one leads to reduced processing delay 
of a packet~\cite{jose2015compiling}. The second one leads to an overall reduction in resource usage and power 
consumption by a V1Model switch. 
Finding  a mapping that has optimal resource usage and does not violate the upper mentioned constraints is 
a computationally intractable problem~\cite{jose2015compiling,vass2020compiling}.

\subsubsection{\textbf{Our approach}}\label{OurApproach}
The TDG mapping  problem can be modeled as an 
\textit{integer linear programming}~\cite{jose2015compiling} problem, 
and an optimal mapping can be 
determined (if it exists). However, it may take a large amount of time~\cite{jose2015compiling,vass2020compiling} 
to compute such a mapping. 
Hence we applied heuristic-based algorithms to compute the mapping. The algorithm works as follows:
First, the algorithm preprocesses (sec.~\ref{TDG:Preprocessing}) the TDG to transform the conditional nodes into MAT. Then it 
loads the TDG into a graph-based data structure with dependencies among the nodes as edge. 
It checks whether the \textit{stateful memory dependencies} in the TDG can be met 
on the target hardware or not. It also transforms the logical MATs to meet the target hardware constraints if necessary. 
Finally, the algorithm maps the logical MATs to the physical match-action stages. 

If a dependency chain in the TDG contains $n$ logical MATs, it requires at least as many match-action stages in the V1Model pipeline. 
Similar to~\cite{jose2015compiling}, we define the \textit{level} of a logical MAT  as the number of \textit{dependencies} remaining 
in any \textit{dependency chain} starting from that node. Assigning the \textit{level}s requires a topological 
ordering of the logical MATs in the TDG. 
Due to topological ordering, all logical MATs at the same level have no dependency on each other. 
Then the mapping algorithm tries to map all the logical MATs of the same  \textit{level}. This process is repeated for all  
\textit{level}s.


\begin{figure}[h]
  \centering
  \includegraphics[trim=0.0in 0in 0in 0, clip,scale=.34]{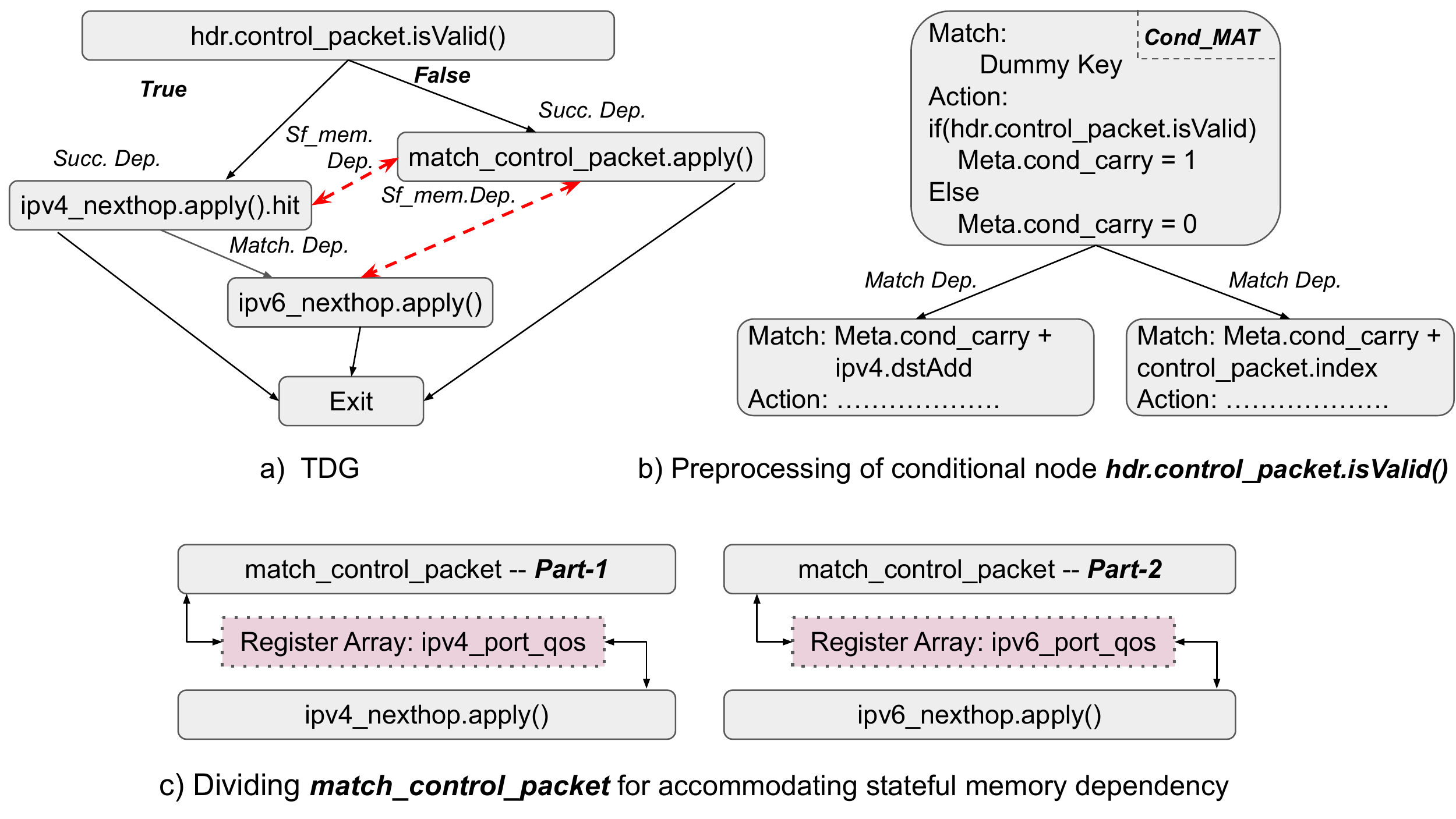}
  \caption{Preprocessing conditional branching and stateful memory access in P4; \textit{Sd\_mem. Dep. = Stateful Memory Dependency}}
  \label{fig:TDGPreprocessing}
\end{figure}

\paragraph{TDG preprocessing}\label{TDG:Preprocessing}

The TDG generated by the P4C frontend transforms every 
conditional branching instruction (if/if-else pair) in the P4 program to 
three logical MATs. 
Fig.~\ref{fig:TDGPreprocessing}a shows the TDG of the P4 code presented in appendix~\ref{App:QoSModiferP4Program}. 
The topmost node shows the TDG node for evaluating the conditional expression, and the next two nodes show its two branches
(one each
for true and false) as the next nodes. However, the task for
evaluating the conditional expression is executed in the action
portion of the first MAT, and it does not pass the result of the expression evaluation to 
the next nodes. Hence, a packet can  successfully match
with both the next nodes, which is incorrect. To avoid this,
we modify the conditional evaluation expression step with 
a conditional assignment available~\cite{bosshart2013forwarding,opentofino} in 
V1Model switches. It  assigns 1 to an auxiliary header
field if the expression evaluates true and 0 if
false (fig.~\ref{fig:TDGPreprocessing}b). This field is added to the match fields of the next nodes; 
both of these MATs are executed for a packet and successfully matche with only one of
them.



\textit{Stateful Memory Dependency} can exist between logical MATs in the same or different paths in the TDG.
Indirect stateful memories accessed by two (or more) logical MATs in the same path implies 
non-stateful memory dependency (sec.~\ref{IR:TDG}) exists among the logical MATs. 
Hence they need to be mapped on different physical match-action stages. 
However,  they need to access the same stateful memory and the RMT architecture does not allow access to same SRAM 
block from differnt match-action stages. 
Therefore, these kinds of 
P4 program can not be implemented using V1Model switches (though the P4 language syntax allows it), and our algorithm rejects them.

On the other hand, logical MATs on  different paths in the TDG with stateful memory dependency  need to be mapped on the same physical 
match-action stages. Hence they need to be assigned the same \textit{level}. 
Besides this, a logical MAT may need to be divided into two separate logical MATs for 
valid mapping. For example, consider the case of three logical MATs ipv4\_nexthop, ipv6\_nexthop, and match\_control\_packet 
in the TDG of fig.~\ref{fig:TDGPreprocessing}a. 
Here match\_control\_packet  is writing over two stateful memories ipv4\_port\_qos, and ipv6\_port\_qos; 
ipv4\_nexthop is reading from only ipv4\_port\_qos and ipv6\_nexthop is reading from only ipv6\_port\_qos.
Now, there exists a \textit{match-dependency} between ipv4\_nexthop and ipv6\_nexthop; hence they can 
not be mapped to same stage. 
Therefore, the stateful memories accessed by these logical MATs also can not be mapped to the same stage. 
Assume ipv4\_port\_qos and ipv6\_port\_qos  are mapped to stage X and Y in the pipeline; 
therefore, ipv4\_nexthop and ipv6\_nexthop are needed to be mapped on stage X and Y, respectively. 
However, this requires match\_control\_packet to be mapped on both stage X and Y; clearly this is not possible. 
To accommodate such scenarios, we bifurcate match\_control\_packet's action into two half based 
on the stateful memory access (fig.~\ref{fig:TDGPreprocessing}c). 
The first half contains the original match and action part up to accessing ipv4\_port\_qos, and 
another logical MAT (match\_control\_packet--Part-2)  is created without any 
match field and only the remaining part of the match\_control\_packet's actions (accessing ipv6\_port\_qos). 
This new logical MAT  is added between match\_control\_packet and its successor in the TDG.

\begin{figure}[h]
\centering
\includegraphics[trim=0.0in 2in 0in 0, clip,scale=.34]{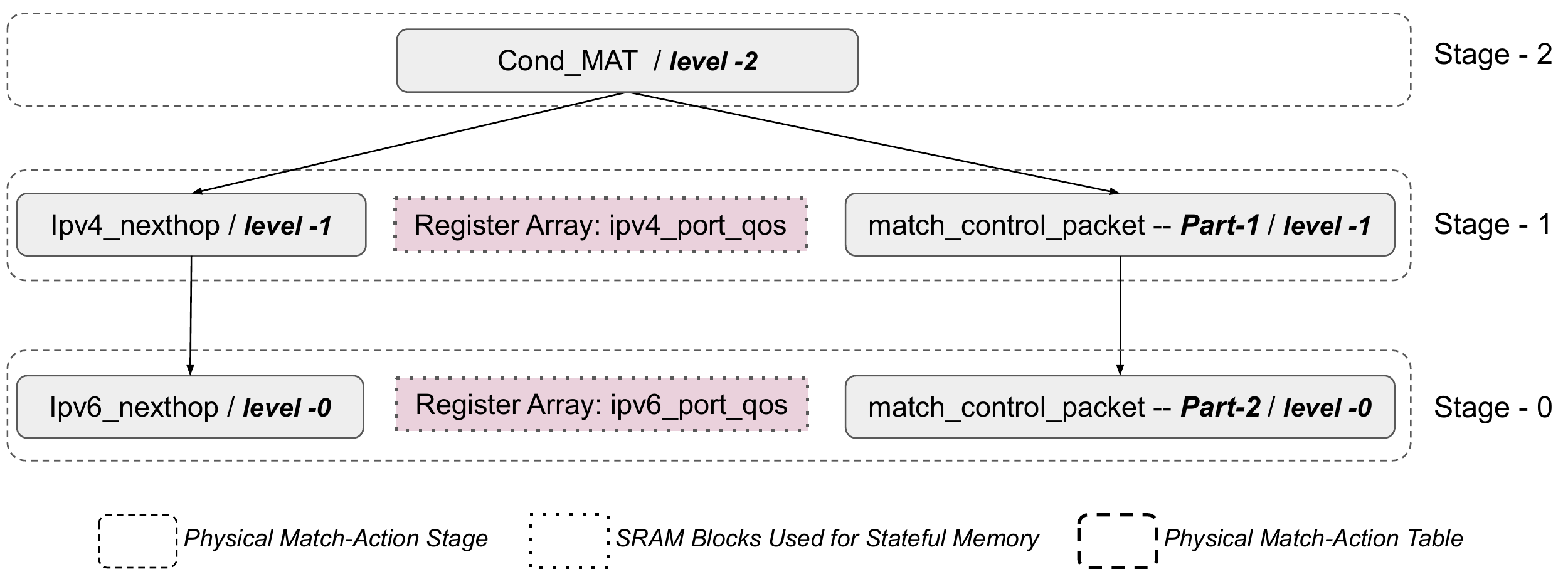}
\caption{The TDG of fig.~\ref{fig:TDGPreprocessing}a after preprocessing, assigned \textit{level} 
for every node and the physical match-action stages after mapping}
\label{fig:LevelAndMapping}
\end{figure}

\paragraph{Level Generation} \label{Mapping:LevelGeneration}
In the TDG, there can be multiple dependencies between logical MATs in the same path; 
however, the strictest dependency (among the four types of non-stateful memory dependencies) 
is  the main factor affecting the mapping decision~\cite{jose2015compiling}. 
The mapping algorithm keeps the \textit{strictest} dependency and removes other dependencies. 
Besides this, all the logical MAT nodes having stateful memory dependency are needed to be 
assigned the same \textit{level} to ensure they are mapped on the same physical match-action stage. 
After the preprocessing step, the TDG nodes are topologically ordered to label them with their appropriate \textit{level}s. 
This ordering ensures that every node with any non-stateful memory dependency is
assigned a higher \textit{level} than its successors. Finally, if two neighbor
nodes in the TDG only have successors or reverse match dependencies
or no dependency, they can be mapped to the same physical match-action stages 
(as they can be executed concurrently or speculatively) and assigned the same \textit{level}.
Fig.~\ref{fig:LevelAndMapping} shows the preprocessed version of the TDG shown in fig.~\ref{fig:TDGPreprocessing},
along with the \textit{level}s assigned to every node.


\paragraph{Mapping Logical Tables}
After \textit{level} generation, all the logical MATs with the same \textit{level}
imply they can be mapped to the same physical match-action stages. They can be divided into two subsets. 
Firstly, a subset of these logical MATs contain stateful memory dependency among them, and 
they need to be mapped on the same stage where the indirect stateful memories are mapped.
Similar to existing commodity RMT switches~\cite{opentofino}, our compiler backend does 
not support spreading indirect stateful memories over multiple stages. 
The mapping algorithm at first maps  the set of logical MATs containing stateful memory 
dependency to only one of the available physical match-action stages. 
Secondly, the following subset of logical MATs contains no match, action, or stateful dependencies, and  
these logical MATs can be executed concurrently on the same math-action stage. 
However, a single match-action stage may not contain enough hardware resources to accommodate them. 
In that case, they are mapped to one or multiple consecutive math-action stages. 

Logical MATs from these two sets can be ordered and selected for mapping to the physical 
match-action stage using various heuristics~\cite{jose2015compiling}. In this work, 
we prioritized the logical MATs with non-exact (ternary, lpm, or prefix) match fields and mapped them to TCAM-based physical match-action blocks. Next, the logical MATs with exact match fields are mapped 
to SRAM-based physical match-action units. These logical MATs can spill into 
TCAMs if it runs out of SRAMs. Ties are broken by at first mapping the logical MAT that appears in the TDG earlier. 
Fig.~\ref{fig:LevelAndMapping} shows the logical MAT nodes of the TDG shown in 
fig.~\ref{fig:TDGPreprocessing} and on which physical match-action stages they are mapped.



In allocating SRAM blocks for the match, action, and stateful memory entries, the mapping algorithm utilizes the 
\textit{memory packing} feature of RMT architecture. The mapping algorithm 
tries to store multiple entries in a \textit{packing unit} of up to $p_f$  SRAM blocks to reduce SRAM waste and fragmentation. 
The value of $p_f$ is configurable at compile time. Currently, our compiler backend allocates SRAM at block level granularity. 
As a result, an SRAM block is allocated exclusively for the match or action entries of only one MAT; or for accomodating the indirect 
stateful memories. It can lead to a waste of SRAM resources when a small number of entries are required. We leave the goal of 
improving the SRAM utilization as future work. 
The number of action entries required for a logical MAT can be determined at compile time~\cite{robin2022clb} or 
unknown beforehand~\cite{jose2015compiling}.  
Our compiler backend can reserve either a fixed number of action entries or 
one action entry for every MAT entry in a logical MAT.

\section{Implementation and Evaluation}  \label{ImplementationAndEvaluation}
\subsection{Implementation} \label{Implementation}
Our P4 compiler backend is entirely implemented in the Python 3 programming language. 
For the frontend, it relies on the P4 consortium's reference compiler implementation (P4C~\cite{P4C}),
which parses the P4 source code and generates an intermediate representation in JSON format. 
Our backend parses  this JSON format data
and stores it in a graph-based data structure. For computing the \textit{parse graph} to \textit{state table}
representation for the parser TCAM,
we relied on the algorithms proposed in~\cite{gibb2013design}. 
This project's source code was written in Python 2 language, and it was not designed for P4\textsubscript{16}'s 
intermediate representation of P4C. We converted the source code in 
Python 3 language and integrated it with our project with a moderate level of modifications. 
All source code of the project is publicly available~\cite{P4CB} under an open-source license.







\subsection{Evaluation} \label{Evaluation}

In this section, we analyze the performance of the proposed P4 compiler backend presented in this work. 
Due to various factors, evaluating and comparing its performance with other compiler backends is challenging. 
A few of the important reasons are the following:
a) Due to the lack of openly available complete compiler backend (which can compute all three types of mapping for a P4 program) 
for RMT switches, we can not compare the overall performance of our compiler backend. 
b) To the best of our knowledge, there are no benchmark p4 programs and complete information about their resource consumption 
in the RMT hardware pipeline is available. 
Few research works have mentioned corresponding P4 program's 
resource consumption using proprietary compilers. However, these target hardware often contains
various externs capable of executing complex actions. These externs can significantly impact the 
mapping of a P4 program. Hence,  
it is not appropriate to use them for comparison without knowing the details 
of the target hardware and relevant mapping algorithm. 
Moreover, a large number of research works are based on the older version of the P4 language (P4\textsubscript{14}). 
P4C can compile both P4\textsubscript{14}  and 
various P4\textsubscript{16} programs targeted for V1Model switches. However, we have found several 
instances~\cite{dang2016paxos,sivaraman2015dc} of 
P4\textsubscript{14} programs used in literature are not compilable using P4C due to a lack of proper backward compatibility. 
Besides this, we have found that several P4\textsubscript{16} 
programs~\cite{dang2016paxos,ding2019estimating} are also
not compilable with P4C due to changes in some of the APIs in BMV2~\cite{BMV2} implementation.
To this end, we have selected the following four P4 programs to evaluate  our compiler backend's performance.

\textit{a) IPv4/IPv6 QoS modifier}: This program is discussed in~\ref{MappingProblem} and the P4 source code is presented 
in appendix~\ref{App:QoSModiferP4Program}.

\textit{b) Simple layer-2/3 forwarding}: This P4 program is designed for simple layer-2/3 packet forwarding
and written using P4 version 14. It is compatible to P4\textsubscript{16}  and used in the TDG mapper proposed in~~\cite{jose2015compiling}. 

\textit{c) Complex layer-2/3 forwarding}: This P4 program is a complex version of the previous one  and requires more resources. 
It is also written using P4 version 14. 
It is compatible to P4\textsubscript{16}  and used in the TDG mapper proposed in~~\cite{jose2015compiling} also. Both versions of the 
layer 2/3 forwarding program was adopted after minor modification to make it compatible with P4C. Moreover, TDG mapping of this P4 program 
is mainly influenced by the availability of memory (SRAM and TCAM) is the pipeline stages. 
Hence, it can be considered as a  memory intensive P4 program.

\textit{d) Traffic Aanonymizer}:  This program is an implementation of the scheme 
proposed in~\cite{kim2019ontas}  to anonymize a packet's content. 
The program is fully realizable using Tofino~\cite{tofino2} switches 
and  BMV2 based source code is  available as open-source project. This program has a small 
stateful memory (SRAM and TCAM) requirement; however, it has a complex TDG and mainly requires computational power in various stages. 
Hence, it can be considered as a compute intensive P4 program. 

The first one is selected to
show some important features of our compiler backend; the next
two are selected as their resource consumption is available in
the literature~\cite{jose2015compiling}. The last one is selected as it is
reported to be realizable using Tofino~\cite{tofino2,opentofino} switches.


For each P4 program, we generated the intermediate representation (IR) using the P4C frontend. 
We used the V1Model switch described in appendix~\ref{APP:ExampleV1ModelSpecs} as the benchmark hardware. 
We provided its hardware specification  and the IR to the compiler backend as input.
Then we used our compiler backend to map the IR of the P4 programs to this benchmark hardware. 
As there is no open-source compiler available for computing the header and parser mapping of a P4 program, we have only presented 
the result using our compiler backend. However, for TDG mapping, we have computed the mapping using the TDG mapper 
presented in~\cite{jose2015compiling} (with \textit{First-Fit-Decreasing} heuristic) and 
compared it with the mapping generated by our backend. The TDG mapper's~\cite{jose2015compiling} mapping algrotihm 
reserves a fixed number of SRAM 
blocks (16 blocks per stage~\cite{jose2015compiling}) for accomodating action memories and can not dynamically adjust the 
number of action entries (for one or more logical MATs) beyond the limit of  16 SRAM blocks. 
As it reserves a fixed number of SRAM blocks, it does not require to compute the mappings of the action entries over the physical MAT stages. 
Besdies the heuristic based mapping, this TDG mapper~\cite{jose2015compiling} can produce optimal mapping of the logical MATs to physical MATs;
However, for optimal mapping it requires a large amount of time~\cite{jose2015compiling,vass2020compiling} to compute the mapping using 
inteleger linear programming (ILP) method. As our goal is to build a compiler backend for to quickly decide about realizability of a P4 program, 
we have not included the ILP variation of the TDG mapper proposed in~\cite{jose2015compiling}.
Compared to the TDG mapper of~\cite{jose2015compiling}, we have configured our compiler backend to allocate upto 16K 
action entries for every logical MAT in every stage. 
All of our experiments were run on  an HP laptop with an Intel Core-i7 processor,
24 GB RAM, running Ubuntu 20.04.

\begin{table*}[t]

\centering{
    \small{
  \begin{tabular}{|c|c|c|c|c|c|}
  \hline
  \begin{tabular}[c]{@{}c@{}}Program  Name\end{tabular} & \begin{tabular}[c]{@{}c@{}}\# Header  Fields\end{tabular} & \begin{tabular}[c]{@{}c@{}}$\sum$ Bitwidth  of Header Fields\end{tabular} & \begin{tabular}[c]{@{}c@{}}$\sum$ Bitwidth of Req. PHV Fields\end{tabular} & \begin{tabular}[c]{@{}c@{}}Waste (\%)\end{tabular} & Ex. Time (in ms) \\ \hline
  QoS-Modifer                                             & 66                                                         & 1288                                                                        & 1432                                                                    & 10.05                                              &2   \\ \hline
  L2L3-Simple                                             & 58                                                         & 1064                                                                        & 1208                                                                    & 11.92                                              &2.03\\ \hline
  L2L3-Complex                                           & 126                                                        & 2912                                                                       & 3088                                                                    & 5.69                                                 &2.12 \\ \hline
  Traffic-Anony                                           & 94                                                         & 1976                                                                       & 2112                                                                    & 6.43                                                &2.92\\ \hline
  \end{tabular}
}
}

\caption{Total number of header fields used in P4 programs (both ingress end egress stage), total bitwidth of the header fields, 
total bitwidth of required PHV fields, percentage of waste in PHV fields and the total execution time 
required for computing the \textit{\textbf{header mapping}}.}
\label{tab::headercomparison}
\end{table*}

\subsubsection{Result Analysis}\label{ResultAnalysis}

\textbf{Header Mapping}: Table~\ref{tab::headercomparison} shows the result of our compiler backend's 
\textit{header mapping} of the benchmark P4 programs. There are 64, 96, and 64 fields of 
8, 16, and 32b width (total 4096b wide PHV) exists in the PHV. 
Many PHV fields remain unused for both small and 
large programs; for example, the small programs (\textit{QoS-Modifier} and \textit{L2L3-Simple}) both consume around 30-35\% 
of the PHV's capacity. On the other hand, the large programs (\textit{Traffic-Anony} and \textit{L2L3-Complex}) 
consume around 51\% of the PHV's capacity. While accommodating the P4 program's header fields  using the PHV fields, some space in the 
PHV fields is wasted. For the small programs, the \textit{waste} in PHV bitwidth
is higher (approx. 10-12\%) compared to 
the larger programs (approx. 5.7-6.4\%). The rightmost column of table~\ref{tab::headercomparison} shows the total time (in milliseconds)
required to compute the \textit{header mapping} for the benchmark P4 programs. For all the P4 programs, the required time is short and ranges 
between 2 to 3 milliseconds (approx.).

\begin{table*}[]
\centering
{
    \small{
\begin{tabular}{|c|c|c|c|c|}
\hline
\begin{tabular}[c]{@{}c@{}}Program  Name\end{tabular} & \begin{tabular}[c]{@{}c@{}}\# States in  Parse Graph\end{tabular} & \begin{tabular}[c]{@{}c@{}}\# Edges in  Parse Graph\end{tabular} & \begin{tabular}[c]{@{}c@{}}Required TCAM Entries\end{tabular} & Ex. time (in ms) \\ \hline
QoS-Modifer                                   & 5                                                                   & 8                                                                 & 5                                                                 & 31 \\ \hline
L2L3-Simple                                             & 4                                                                   & 10                                                                 & 5                                                      & 21         \\ \hline
L2L3-Complex                                           & 11                                                                  & 31                                                                & 22                                                       &  132        \\ \hline
Traffic-Anony                                           & 7                                                                   & 14                                                                & 13                                                      & 65        \\ \hline
\end{tabular}
}
}
\caption{
  Total number of states and edges in the parse graph of the
  P4 programs, number of TCAM entries  required for the \textit{state table} and 
  the total execution time required for computing the \textit{\textbf{parse graph mapping}}.}
\label{tab::parserComparison}  
\end{table*}

\textbf{Parse Graph Mapping}: 
Table~\ref{tab::parserComparison} shows the result of our compiler backend's 
\textit{header mapping} of the benchmark P4 programs. The benchmark hardware used in the 
experiments contains 256$\times$40b TCAM for implementing the parser \textit{state table};
it can look at 48 bytes into the packet, identify a maximum of four headers and extract 48 bytes of header fields data at every cycle to parse 
the incoming packets at  40 Gbps. For the complex P4 program (\textit{L2L3-Complex}) with 11 states and 
31 edges in the parse graph, it requires only 22 entries in the TCAM, which is less than 9\% of the total capacity of the TCAM. 
The \textit{parse graph} is simpler for the
rest of the programs and consumes only 2\% of the TCAM's capacity. 
The rightmost column of table~\ref{tab::parserComparison} shows the total time (in milliseconds)
required to compute the \textit{parse graph mapping} for the benchmark P4 programs. 
The result shows
that for the P4 programs with a larger number of nodes and edges in the \textit{parse graph}, 
the \textit{parse graph mapping} algorithm requires more time to compute the mapping. 
For example, the total number of nodes and edges in the \textit{parse graph} of \textit{L2L3-Complex} is approximately  twice
as compared to  the \textit{parse graph} of the \textit{Traffic-Anony} program. 
Computing the \textit{parse graph mapping} for \textit{L2L3-Complex} program 
requires approximately 2x time compared to the  \textit{Traffic-Anony} program.
In the other two programs, the number of nodes and edges in
the \textit{parse graph} is small; hence 
the execution time of the \textit{parse graph mapping} is also relatively short (less than 35 milliseconds).

\begin{table*}[!t]
\centering{
  \begin{tabular}{|c|c|c|cc|cc|cccc|cc|}
    \hline
    \multirow{3}{*}{\textbf{Program Name}} & \multicolumn{1}{l|}{\multirow{3}{*}{\# Nodes in TDG}} & \multicolumn{1}{l|}{\multirow{3}{*}{\# Edges in TDG}} & \multicolumn{2}{c|}{\multirow{2}{*}{Stages}} & \multicolumn{2}{c|}{\multirow{2}{*}{Latency (in cycle)}} & \multicolumn{4}{c|}{Resource Usage  (in Blocks)}                                                                   & \multicolumn{2}{c|}{\multirow{2}{*}{Ex. Time (in ms)}} \\ \cline{8-11}
                                           & \multicolumn{1}{l|}{}                                 & \multicolumn{1}{l|}{}                                 & \multicolumn{2}{c|}{}                        & \multicolumn{2}{c|}{}                                    & \multicolumn{2}{c|}{TCAM}                                          & \multicolumn{2}{c|}{SRAM}                     & \multicolumn{2}{c|}{}                                  \\ \cline{4-13} 
                                           & \multicolumn{1}{l|}{}                                 & \multicolumn{1}{l|}{}                                 & \multicolumn{1}{c|}{~\cite{jose2015compiling}} & *  & \multicolumn{1}{c|}{~\cite{jose2015compiling}}       & *        & \multicolumn{1}{c|}{~\cite{jose2015compiling}} & \multicolumn{1}{c|}{*}   & \multicolumn{1}{c|}{~\cite{jose2015compiling}} & *   & \multicolumn{1}{c|}{~\cite{jose2015compiling}}     & *        \\ \hline
    QoS-Modifer                            & 16                                                    & 20                                                    & \multicolumn{1}{c|}{Inv.}               & 3  & \multicolumn{1}{c|}{Inv.}                     & 38       & \multicolumn{1}{c|}{Inv.}               & \multicolumn{1}{c|}{6}   & \multicolumn{1}{c|}{Inv.}               & 4   & \multicolumn{1}{c|}{Inv.}                   & 31       \\ \hline
    Traffic-anony                          & 84                                                    & 194                                                   & \multicolumn{1}{c|}{Inv.}               & 18 & \multicolumn{1}{c|}{Inv.}                     & 156      & \multicolumn{1}{c|}{Inv.}               & \multicolumn{1}{c|}{2}   & \multicolumn{1}{c|}{Inv.}               & 35  & \multicolumn{1}{c|}{Inv.}                   & 1700     \\ \hline
    L2L3-simple                            & 24                                                    & 38                                                    & \multicolumn{1}{c|}{4}                  & 5  & \multicolumn{1}{c|}{53}                       & 42       & \multicolumn{1}{c|}{56}                 & \multicolumn{1}{c|}{57}  & \multicolumn{1}{c|}{191}                & 207 & \multicolumn{1}{c|}{243}                    & 41       \\ \hline
    L2L3-Complex                           & 60                                                    & 138                                                   & \multicolumn{1}{c|}{30}                 & 31 & \multicolumn{1}{c|}{110}                      & 108      & \multicolumn{1}{c|}{272}                & \multicolumn{1}{c|}{260} & \multicolumn{1}{c|}{762}                & 995 & \multicolumn{1}{c|}{148}                    & 925      \\ \hline
    \end{tabular}
}
    \caption{Comarison of \textit{\textbf{TDG mapping}} computed by the work proposed in~\cite{jose2015compiling} and the compiler 
backend presented in this work (\textit{marked by *}): Inv. = Invalid mapping.}
\label{tab::TDGmapperComparison} 
\end{table*}

\textbf{TDG Mapping}:
Table~\ref{tab::TDGmapperComparison}
shows the comparison between the TDG mapping computed by the TDG mapper of~\cite{jose2015compiling} and our compiler backend. 
For the \textit{QoS-Modifier} program, the TDG mapper proposed in~\cite{jose2015compiling} computes the mapping wrongly. 
It maps \textit{match\_control\_packet} and \textit{ipv4\_nexthop} in one stage and \textit{ipv6\_nexthop} in another stage. 
This is an invalid mapping, as explained in sec.~\ref{OurApproach} and fig.~\ref{fig:TDGPreprocessing}.
The reason behind the invalid mapping (row 1 in table~\ref{tab::TDGmapperComparison}) is that the 
TDG mapper proposed in~\cite{jose2015compiling} only focuses on 
the four types of  
non-stateful memory dependencies (sec.~\ref{IR:TDG}) in computing the mapping. 
However, the \textit{QoS-Modifier} program contains a sequence of 
logical MATs where both match and stateful memory dependency guide the control flow.
Hence, the mapper fails to compute a valid mapping for the \textit{QoS-Modifier} program. On the other hand, 
our compiler backend includes stateful memory dependency and the other four types of non-stateful memory dependencies in 
computing the TDG mapping.
 It generates a valid mapping for this P4 program. It maps the logical MATs over three physical match-action stages, and the processing 
latency of every packet under this mapping is 38 cycles.  
It requires 6 TCAM blocks and 4 SRAM blocks to accommodate the logical match-action tables.

The \textit{Traffic-Anony} program is written in P4\textsubscript{16}. Whereas the work described in~\cite{jose2015compiling}
does not support various P4\textsubscript{16} language constructs and 
only works with the P4\textsubscript{14} programs. Hence it can not generate the mapping for 
 the \textit{Traffic-Anony} P4 program (row 2 in table~\ref{tab::TDGmapperComparison}). 
However, our compiler backend supports the  P4\textsubscript{16} language and generates a valid mapping for the program. 
It maps the P4 program over 18 physical match-action stages and the processing 
latency of every packet under this mapping is 156 cycles.  
The \textit{Traffic-Anony}  is a computation-intensive  P4 program; it 
requires 2 TCAM blocks and 35 SRAM blocks to accommodate the logical match-action tables in the RMT pipeline.

The P4 programs for simple and complex L2L3 forwarding (\textit{L2L3-Simple} and \textit{L2L3-Complex}) 
are both written in P4\textsubscript{14}. None of the programs contain any \textit{stateful memory dependency}~\ref{TDG:Preprocessing}.  
Hence their TDG mapping can be computed using the work proposed in~\cite{jose2015compiling}.
In the case of \textit{L2L3-Simple} (row 3 in table~\ref{tab::TDGmapperComparison}), 
our compiler backend uses five stages but achieves reduced packet processing latency of 42 cycles. 
On the other hand, the TDG mapper of~\cite{jose2015compiling} uses one less stage, but the packets face more
processing latency (53 cycles). Our compiler backend requires one extra TCAM block and 16 SRAM blocks to 
accommodate the logical MATs.
In the case of \textit{L2L3-Complex}, our compiler backend uses 31 stages which is one stage more compared to~\cite{jose2015compiling}. 
However, it needs a smaller packet processing latency of 108 cycles. 
Our compiler backend uses 260 TCAM blocks compared to 272 TCAM blocks used by the TDG mapper of~\cite{jose2015compiling}. 
However, our compiler backend uses a higher number of SRAM blocks (995 blocks compared to 762 blocks).  
Overall our compiler backend requires more SRAM blocks for accommodating the simple and complex L2L3 forwarding programs. 
There are two reasons behind this: Firstly, our compiler backend  allocates SRAM blocks for action memory and indirect stateful memories 
at a granularity of blocks. As a result, when 
the SRAM requirement of two or more logical MATs can be fulfilled using only one SRAM block, our compiler backend 
allocates at least one SRAM block to every logical MAT. This memory packing is less efficient compared to~\cite{jose2015compiling}. 
 We leave the goal of improving the SRAM allocation mechanism  to reduce such waste as a 
future goal. Finally, our compiler backend does not reserve a fixed number of SRAM blocks for action memories in every stage. 
Instead, it can store up to 16K (can be increased or decreased at compile time) action entries for every logical MAT. 
It sacrifices cheaper SRAM uses for more flexibility in action memory allocation.  
As a result, it tends to use more SRAM blocks compared to the TDG mapper of~\cite{jose2015compiling}.

\textit{Execution time}: The TDG mapper of~\cite{jose2015compiling} can not compute a valid  TDG mapping for 
the \textit{QoS-Modifier} and \textit{Traffic-Anony} program; hence execution time for these two benchmark P4 programs 
can not be compared.  However, in the case of  \textit{L2L3-Complex}, our compiler backend 
requires approximately 6x time  compared to the TDG mapper of~\cite{jose2015compiling}
(column 12-13, last/5-th row in table~\ref{tab::TDGmapperComparison})
to compute the \textit{TDG mapping}. There are two major reasons behind this:
Firstly, our compiler backend  preprocesses (sec.~\ref{TDG:Preprocessing}) the TDG to handle 
the \textit{stateful memory dependency}, which increases the execution time. 
Secondly, the TDG mapper's~\cite{jose2015compiling} policy of allocating a  fixed number of SRAM blocks per stage for action memories
does not require any computation at run time. Conversely, our compiler backend dynamically allocates action
memories for every logical table and also tries to minimize
the usage of SRAM blocks. 
Hence it needs to search over a larger
search space which results in a larger execution time.
The \textit{L2L3-Complex} is a memory-intensive P4 program; it contains 60 nodes and 138 edges in the TDG. 
Here the logical tables
need a large number of matches and action entires. As a result,
our compiler backend needs to do more computation due to
the mentioned two factors and requires more execution time (925 ms compared to 148 ms). 
On the other hand, for the \textit{L2L3-Simple} program (4-th row in table~\ref{tab::TDGmapperComparison}), 
the TDG is less complex (24 nodes and 38 edges); its logical tables also requires a smaller number of match and action entries. 
Hence,
our compiler backend quickly computes  the TDG mapping 
and requires approximately 1/6-th time (41 ms vs 243 ms) compared to the TDG
mapper of [17].

Our compiler backend's larger execution time can be observed in the case of the \textit{Traffic-Anony} program also.
It is a  computation-intensive P4 program (3-rd row in table~\ref{tab::TDGmapperComparison}) containing a large  number of 
nodes and edges in the TDG (84 nodes and 194 edges) due to the nested branching instructions. However, it 
requires very little memory (only 2 TCAM and 35 SRAM blocks). Hence, the \textit{TDG mapping}  of this program 
is mainly influenced by the available bit width in the match ($TCB$ and $SCB$ in fig.~\ref{fig:RMTSingleStage}) 
and action ($ACB$ in fig.~\ref{fig:RMTSingleStage}) crossbars. The use of crossbar widths is 
decided by the use of various PHV fields  in the match and action portion of a physical MAT. 
The benchmark hardware used in the  evaluation (app.~\ref{APP:ExampleV1ModelSpecs}) contains 226 PHV fields of 4 different bitwidths. 
As a result, our compiler backend needs to select a good mapping from a large number of variations, and the execution time increases (1700 ms). 

Though our compiler backend needs more execution time compared to heuristic-based  algorithms of the 
TDG mapper of~\cite{jose2015compiling}; its execution time 
is very short compared to the integer linear programming (ILP) based optimal mapping algorithms. 
For example, the ILP based optimal mapping algorithms require 12K-135K ms (table 4 in~\cite{jose2015compiling}) execution time 
for the \textit{L2L3-Complex} program compared to sub 1K ms execution time of our compiler backend. 
Moreover, our compiler backend is also capable of dynamically allocating action memories for every logical MAT. 
Besides the statistics shown in table~\ref{tab::TDGmapperComparison}, our compiler backend also provides details information 
about which logical MATs are mapped to which physical match-action stages. 
It also provides the required number of TCAM and(or) SRAM blocks, match and action 
crossbar width,  type of ALU/extern instructions, etc., required for every logical MAT. This information is 
required in the \textit{Configuration Generation Phase} of a compiler backend. 
As this phase is not the focus of this work, we do not discuss them here in detail. However, more details are provided in
our GitHub repository~\cite{P4CB}. 
Overall, our compiler backend
provides a feature-rich open source alternative to the commercial closed source compilers that can quickly decide 
the realizability of a P416 program.


\section{Discussion}  \label{Discussion}

\textbf{Limitations}:
Our compiler backend supports most of the P4 language constructs covering a wide range of use cases. 
However, it still does not support variable-length header parsing and direct stateful memory access in actions. 
Both of them can be avoided through careful design of the p4 program. 
Besides this, it does not support the atomic transaction mechanism available in the P4 language. 
We are working on supporting these  P4 language features.

\textbf{Extending  V1Model Architecture}:
The PSA~\cite{PSA} or Tofino~\cite{tofino2} is an extension of V1Model architecture where the architectures support different externs. 
These architectures can combine multiple simpler instructions in one atomic instruction to achieve complex functionalities. 
For example, the \textit{register extern} available in Tofino switches~\cite{opentofino} can execute 
four-way branching instructions. It can execute two if-else pairs and read-modify-write operation on a pair of 
registers (indirect stateful memory) using only one extern. 
However, to use them (or any new extern in general) in a P4 program, the P4C compiler frontend needs to support them. 
After that, these externs can be 
supported in our compiler backend with minor modifications to compute mapping for that P4 program. 

\textbf{Writing New Mapping Algorithm}:
Our compiler backend is designed in a modular way. After parsing the intermediate representation of a P4 program, 
it stores the preprocessed information (header information, parse graph, TDG) in various convenient data 
structures (hash table, graph, etc.). 
Besides this, it also stores the resources in a  V1Model  switch in various convenient data structures 
(hashtable, array, etc.). As an open-source project, researchers can reuse this processed information to 
write newer algorithms for header mapping, parse graph mapping, and TDG mapping. 
Detail discussion on source code organization is available in~\cite{P4CB}.




\section{Conclusion}  \label{Conclusion}
We have presented an open-source compiler backend that can map a P4 (version 16) program to the hardware resources 
of a  V1Model switch. 
It uses heuristic-based algorithms to compute this mapping and give a quick decision on the realizability of a P4 program. 
We believe this open-source compiler backend can serve as a cost-effective platform for analyzing the realizability 
and resource consumption 
of a P4 (version 16) program in real-world V1Model  switches. It allows researchers to experiment with different mapping algorithms as an open-source platform. 
It can be extended to support other derivatives of V1Model architecture by supporting various \textit{extern} units. 
This can provide 
an open platform to the programmable switch researchers for experimenting with different mapping algorithms and different 
variations of V1Model switch. 

\bibliographystyle{unsrt}
\bibliography{Compiler}


\appendix
\section{\textit{QoS-Modifer} P4 Program}\label{App:QoSModiferP4Program}
The P4\textsubscript{16} source code for the \textit{QoS-Modifier} program. 
We have implemented the program in P4\textsubscript{14} language and used it in sec.~\ref{Evaluation}. 
The P4\textsubscript{14} version of the program is compatible with the TDG mapper described in~\cite{jose2015compiling}.

\small{
\begin{lstlisting}[linewidth=\columnwidth,breaklines=true,frame = single]
{
header control_packet_t {
bit<8> ipv4_diffserv;
bit<8> ipv6_trafficClass;
bit<8> index;
}
header ipv6_t {
bit<4>   version;
bit<8>   trafficClass;
bit<20>  flowLabel;
bit<16>  payloadLen;
bit<8>   nextHdr;
bit<8>   hopLimit;
bit<128> srcAddr;
bit<128> dstAddr;
}
header ethernet_t {
bit<48> dstAddr;
bit<48> srcAddr;
bit<16> etherType;
}
header ipv4_t {
bit<4>  version;
bit<4>  ihl;
bit<8>  diffserv;
bit<16> totalLen;
bit<16> identification;
bit<3>  flags;
bit<13> fragOffset;
bit<8>  ttl;
bit<8>  protocol;
bit<16>  
hdrChecksum;
bit<32> srcAddr;
bit<32> dstAddr;
}
struct headers {
Control_packet_t control_packet;
Ethernet_t ethernet;
ipv4_t ipv4;
Ipv6_t ipv6;
}

// Parsing logic starts from here

state parse_ethernet {
// Extract 14 bytes ethernet header & transition to next states
packet.extract(hdr.ethernet);
  transition select(hdr.ethernet.etherType){
      16w0x800: parse_ipv4;
      16w0x86dd: parse_ipv6;
      16w0x806: parse_control;
      default: accept;
  }
}
state parse_control {
// Extract 3 bytes control_packet //header
packet.extract(hdr.control_packet);
transition accept;
}
state parse_ipv4 {
// Extract 20 bytes IPv4 header
packet.extract(hdr.ipv4);
transition accept;
}
state parse_ipv6 {
//Extract 40 bytes IPv6 header
packet.extract(hdr.ipv6);
transition accept;
}
state start {
transition parse_ethernet;
}


// Definition of the stateful memory (register) array 
register<bit<8>, bit<7>>(32w128) ipv4_port_qos;
register<bit<8>, bit<7>>(32w128) ipv6_port_qos;

// Table Definition starts from here
action set_ipv4_ipv6_qos() {
ipv4_port_qos.write((bit<32>)hdr.control_packet.index, hdr.control_packet.ipv4_diffserv);                                                  
ipv6_port_qos.write((bit<32>)hdr.control_packet.index, hdr.control_packet.ipv6_trafficClass);              
}
table match_control_packet {
actions = {
  set_ipv4_ipv6_qos;
}
key = {
  hdr.control_packet.index: exact;
}
size = 256;
}
action set_next_hop_ipv4(bit<9>port) {
meta.egress_port = port;
ipv4_port_qos.read(
hdr.ipv4.diffserv,(bit<32>)port);
hdr.ipv6.dstAddr = SERVER_IP;
}
table ipv4_nexthop {
actions = {
  set_next_hop_ipv4;
  ipv4_miss;
}
key = {
  hdr.ipv4.dstAddr: exact;
}
size = 256;
}
action set_next_hop_ipv6(bit<9> port) {  
meta.egress_port = port;
ipv6_port_qos.read(
hdr.ipv6.trafficClass,  (bit<32>)port);
}
ipv6_nexthop {
actions = {
  set_next_hop_ipv6;
}
key = {
  hdr.ipv6.dstAddr: exact;
}
size = 256;
}

/// Control Flow starts from here
control ingress(......) {
apply {
  if (hdr.control_packet.isValid()) {
    match_control_packet.apply();
  } 
  else {
    ipv4_nexthop.apply();
    if(ipv4_nexthop.hit){
        ipv6_nexthop.apply();
    }
  }
}
}
}
\end{lstlisting}
}

\section{Example V1Model Hardware Sepcification} \label{APP:ExampleV1ModelSpecs}

Following JSON instance is written according to the \textit{Hardware Specification Language} presented 
in sec~\ref{HSLSection}. All the examples presented in sec~\ref{MappingProblem} and the evaluation 
of our compiler backend presented in sec.~\ref{Evaluation} is based on the V1Model hardware represented 
by this JSON instance.
Attributes in this JSON instance are annotated with the notations 
used to specify the hardware resources 
of V1Model switch in sec.~\ref{V1ModelArchitecture}.


\small{
\begin{lstlisting}[linewidth=\columnwidth,breaklines=true,frame = single,mathescape=true]
{
"Name": "RMTV1Model32Stage",
"TotalStages": 32,
"HeaderVectorSpecs": [
{
  "BitWidth" ($f_w^1$): 8,
  "Count" ($f_C^1$): 64,
  "ALU": "8BitAlu"
},
{
  "BitWidth" ($f_w^2$): 16,
  "Count" ($f_C^2$): 96,
  "ALU": "16BitAlu"
},
{
  "BitWidth" ($f_w^3$): 32,
  "Count" ($f_C^3$: 64,
  "ALU": "32BitAlu"
}
],
"ParserSpecs": {
  "ParsingRate" ($P_{Rate}$): 60,
  "HeaderIdentificationBufferSize" ($P_B$): 48,
  "MaxIdentifieableHeader" ($H$): 4,
  "MaxMoveAheadBit" ($P_{MA}$: 128,
  "TCAMLength" ($P_L^T$): 256,
  "TCAMLookupFieldCount" ($f_C^T$): 4,
  "TCAMLookupFieldWidth" ($f_W^T$): 1,
  "MaxExtractableData" ($P_W^E$): 48
},
"StageDescription": [
{
  "Index": "0-31",
  "PerMATInstructionMemoryCapacity" ($A_C$): 32,
  "ActionCrossbarBitWidth" ($ACB_W$): 1280,
  "SRAMResources": {
    "MemoryPortWidth" ($M_BW$): 80,
    "MemoryPortCount" ($M_P$): 8,
    "MemoryBlockCount" ($S$): 106,
    "MemoryBlockBitWidth"  ($S_W$): 80,
    "MemoroyBlockRowCount"  ($S_L$): 1024
  },
  "TCAMMatResources": {
    "MatchCrossbarBitWidth" ($TCB_W$): 640,
    "BlockCount" ($T$): 16,
    "SupportedMatchTypes": {
      "exact": true,
      "lpm": true,
      "range": true,
      "ternary": true
    },
    "PerTCAMMatBlockSpec": {
      "TCAMBitWidth" ($T_W$): 40,
      "TCAMRowCount" ($T_L$): 2048
    }
  },

  "SRAMMatResources": {
    "MatchCrossbarBitWidth" ($$): 640,
    "BlockCount" ($$): 8,
    "SupportedMatchTypes": {
      "exact": true,
      "lpm": false,
      "range": false,
      "ternary": false
    },
    "PerSRAMMatBlockSpec": {
      "SRAMBitWidth" ($$): 80,
      "HashingWay" ($$): 4
    }
  },
  "ExternResources": {
  "RegisterExtern": [
  {
    "name": "RegisterExtern_8"
  },
  {
    "name": "RegisterExtern_16"
  },
  {
    "name": "RegisterExtern_32"
  },
  {
    "name": "RegisterExtern_64"
  }
],"CounterExtern": [
  {
    "name": "CounterExtern_Packet"
  },
  {
    "name": "CounterExtern_Byte"
  },
  {
    "name": "CounterExtern_PacketByte"
  }
],"MeterExtern": [
  {
    "name": "MeterExtern_Byte"
  },
  {
    "name": "MeterExtern_Packet"
  },
  {
    "name": "RegisterExtern_32"
  }
]
}
}
],
"SingleStageCycleLength": 14,
"DependencyDelayInCycleLegth": {
  "match_dependency": 12,
  "action_dependency": 3,
  "successor_dependency": 1,
  "reverse_match_dependency": 1,
  "default": 1
}
}
\end{lstlisting}

\end{document}

